\documentclass[11pt]{article}
\usepackage{fullpage}
\usepackage{datetime}
\usepackage{amsmath}
\usepackage{amssymb}
\usepackage{mathtools}
\usepackage{graphics}
\usepackage{graphicx}
\usepackage{color}
\usepackage{xspace}
\usepackage{latexsym}
\usepackage{epsf}
\usepackage{multirow}
\usepackage{url}
\usepackage{hyperref}

\newcommand{\AND}{\wedge}
\newcommand{\OR}{\vee}

\newcommand{\comment}[1]{{}}

\newcommand{\when}{\ \mbox{${:\!\!-\;}$}}

\newcommand{\myc}[1]{}

\newcommand{\inst}{\chi}
\newcommand{\noninst}{\overline{\chi}}
\newcommand{\troot}{\pi}
\newcommand{\vars}{\id{vars}}
\newcommand{\abstr}{{\mathtt{abs}}}
\newcommand{\alt}[2]{{#1\!\!:\!\!#2}}

\newtheorem{Lem}{Lemma}

\newtheorem{Thm}[Lem]{Theorem}
\newtheorem{Pro}[Lem]{Proposition}

\newtheorem{Def}{Definition}

\newtheorem{Ex}{Example}

\newcommand{\id}[1]{\mbox{\it #1\/}}

\newcommand{\kw}[1]{\mbox{\tt #1}}

\def\p@enumiii{\theenumi(\theenumii)}

\renewcommand{\subparagraph}[1]{\smallskip \noindent \textbf{\textit{#1}} \hspace*{0.5em}}

\newcommand\annotate[1]%
{\textcolor{blue}{$^\maltese$}%
\-\marginpar[\raggedleft\scriptsize \textcolor{red}{#1}]%
{\raggedright\scriptsize \textcolor{red}{#1}}}

\begin{document}

\title{\textbf{Model Checking with Probabilistic
    Tabled~Logic~Programming}\thanks{%
A prototype implementation of the techniques described in this paper
is available at
\texttt{\href{http://www.cs.stonybrook.edu/~cram/probmc/}{http://www.cs.stonybrook.edu/{\scriptsize
      $\sim$}cram/probmc}}.
}
}

\author{
\begin{minipage}{5in}
\begin{tabular}{c}
{\Large Andrey Gorlin, C.\ R.\ Ramakrishnan, and Scott A.\
  Smolka}\\[1em]
  Department of Computer Science\\
  Stony Brook University,  Stony Brook, NY 11794-4400, U.S.A. \\
  \texttt{\href{mailto:cram@cs.stonybrook.edu}{\{agorlin, cram, sas\}@cs.stonybrook.edu}}
\end{tabular}
\end{minipage}
}

\date{}

\maketitle
\begin{abstract}
We present a formulation of the problem of probabilistic model
checking as one of query evaluation over probabilistic logic programs.
To the best of our knowledge, our formulation is the first of its kind,
and it covers a rich class of probabilistic models and probabilistic
temporal logics.  The inference algorithms of existing probabilistic
logic-programming systems are well defined only for queries with a
finite number of explanations.  This restriction prohibits the encoding
of probabilistic model checkers, where explanations correspond to
executions of the system being model checked.  To overcome this
restriction, we propose a more general inference algorithm that uses
finite generative structures (similar to automata) to represent
families of explanations.  The inference algorithm computes the
probability of a possibly infinite set of explanations directly from
the finite generative structure.
We have implemented our inference algorithm in XSB Prolog, and use this
implementation to encode probabilistic model checkers for
a variety of temporal logics, including PCTL and GPL (which subsumes
PCTL$^*$).  Our experiment results
show that, despite the highly declarative nature of their encodings, the
model checkers constructed in this manner are competitive with their
native implementations.
\end{abstract}

\section{Introduction}
\label{sec:intro}

Beginning in 1997, we formulated the problem of model checking as one
of query evaluation over logic programs~\cite{RRRSSW:CAV97}.  The
attractiveness of this approach is that the operational semantics of
complex process languages (originally CCS~\cite{Mil89}, followed by
value-passing calculi~\cite{Ram:PADL01}, the pi-calculus~\cite{MPW92},
and mobile calculi with local broadcast~\cite{SRS:Coord08}), as well
as the semantics of complex temporal logics (e.g., the modal
mu-calculus~\cite{Kozen83}), can be 
expressed naturally and at a high level as clauses in a logic program.
Model checking over these languages and logics then becomes query
evaluation over the logic programs that directly encode their
semantics.

\comment{
Our work on model checking as query evaluation over logic programs
has exploited various features of the logic-programming
paradigm to advance the state of the art of verification.  Our original
work on CCS and the modal mu-calculus took advantage of \emph{tabled
evaluation} for efficient fixed-point 
computation.  Our encoding for the pi-calculus used \emph{logical
variables} to represent names, thereby using the inference system to
automatically rename them as needed or to introduce fresh names and to
handle the creation and modification of scopes~\cite{YRS:STTT04}.  Our approach
to model checking data-independent systems relied on the use of
constraints and unification to avoid enumerative binding of
variables~\cite{SR:ICFEM03}.  Model checking timed systems in our
framework used constraint handling (over the
reals)~\cite{DRS:RTSS00,PRR:ICLP02}.  Other logic-programming
techniques have played a role in our success, ranging from simple
translation schemes (for compiling transition rules~\cite{DR:FORTE99}) to
fold-unfold transformations (for generating automated induction
proofs~\cite{RKRRS:TACAS00}).\annotate{Cut this para, cite these
  papers in related work to save space?}
}

The past two decades have witnessed a number of important
developments in Probabilistic Logic Programming (PLP), combining
logical and statistical inference, and leading to a number of
increasingly mature PLP implementations.
A natural question is whether the advances in PLP enable the development
of model checkers for \emph{probabilistic systems}, the same way
traditional LP methods such as tabled evaluation and constraint
handling enabled us to formulate model checkers for a variety of
non-probabilistic systems.

It turns out that existing %(combined statistical and logical)
PLP inference methods are not sufficiently powerful to be used
as a basis for probabilistic model checking.  One of the earliest PLP
inference procedures, used in PRISM~\cite{sato-kameya-prism97},
is formulated in terms of the set of
\emph{explanations} of answers.  PRISM puts in place three
restrictions to make its inference work:
(a)~\emph{independence}: random variables used in any single
explanation are all independent; (b)~\emph{mutual exclusion}: two distinct
explanations of a single answer are mutually exclusive; and
(c)~\emph{finiteness}: the number of possible explanations of an answer is
finite.  Subsequent systems, notably ProbLog~\cite{Problog} and
PITA~\cite{RiguzziSwift10a} have eliminated
the independence and mutual exclusion restrictions
of PRISM.  This, however, is still insufficient for model checking, as
the following example shows.

\begin{figure}
  \centering
  \begin{tabular}[c]{cc}
\begin{minipage}[c]{2in}
\includegraphics[width=1.75in]{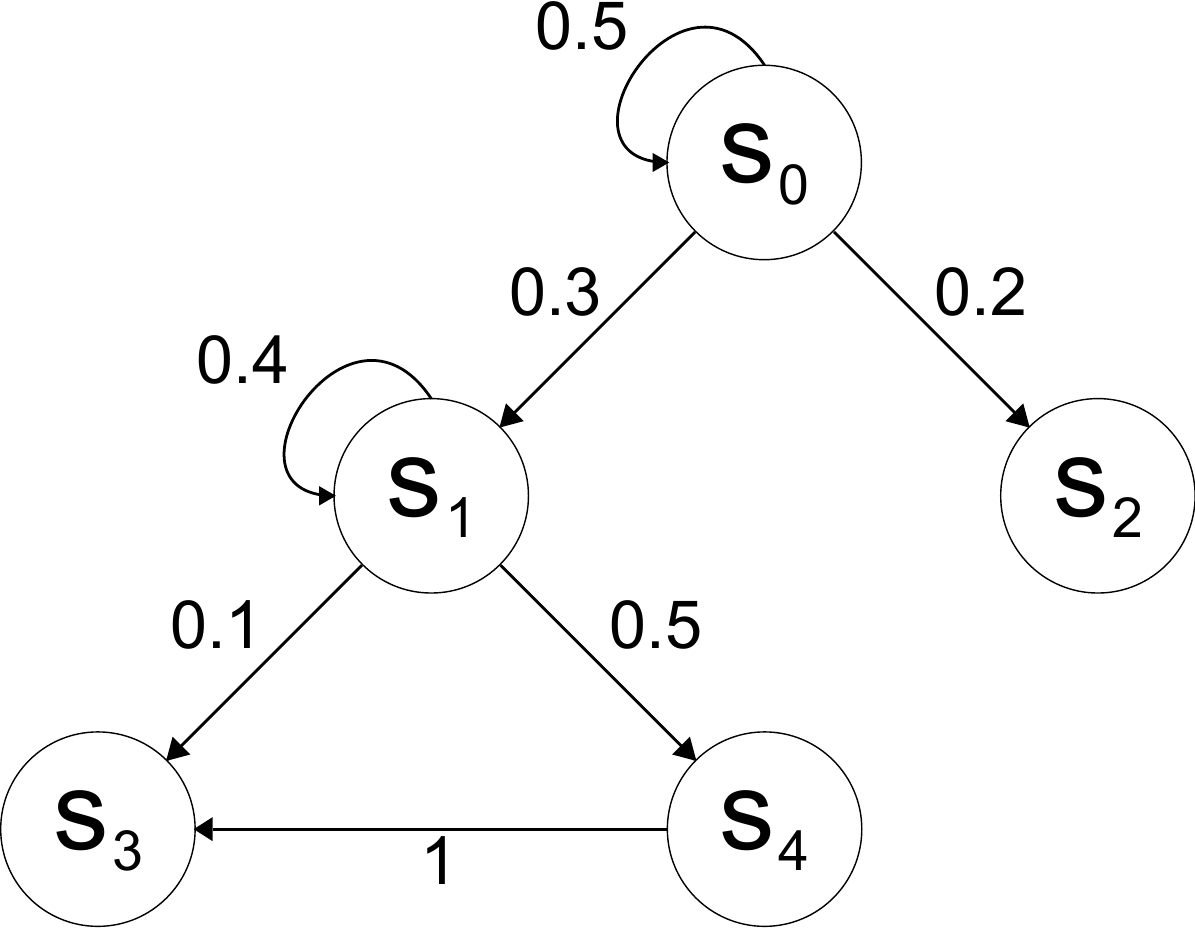}
\end{minipage}
& 
\begin{minipage}[c]{2.75in}
\scriptsize
\begin{verbatim}
% 3 "switches" (random processes) for transitions
% from states s0, s1 and s4, respectively.
values(t(s0), [s0, s1, s2]).
values(t(s1), [s1, s3, s4]).
values(t(s4), [s3]).

% Distribution parameters of the random variables.
set_sw(t(s0), [.5, .3, .2]).
set_sw(t(s1), [.4, .1, .5]).
set_sw(t(s4), [1]).

% Transition from S at instance I goes to T,
% as determined by the corresponding random process.
trans(S, I, T) :-
    msw(t(S), I, T).

% Starting at state S at instance I, state T is reachable.
reach(S, I, T) :-
    trans(S, I, U),
    reach(U, next(I), T).
reach(S, _, S).
\end{verbatim}
\end{minipage}\\
(a) & (b)
  \end{tabular}
 \caption{(a)~Example Markov chain; (b)~PRISM encoding of transitions
   in the chain.}\label{fig:runex}
\end{figure}

\paragraph{Motivating Example:}  Figure~\ref{fig:runex} shows a Markov
chain and its representation in PRISM.  Note that the behavior of a
Markov chain is \emph{memoryless}: in any execution of the
chain, a transition from state, say $s$, is independent of any previous
transitions (including those from the same state).  The definition of
the \texttt{trans} predicate has an explicit instance parameter
\texttt{I}, which is subsequently used in \texttt{msw}.  PRISM
treats different instances of the same random
variable as independent.  Thus \texttt{trans} correctly encodes the
semantics of the Markov chain.
 
We first consider simple reachability questions of the form: What is
the likelihood that on an execution of the chain from a start state
$s$, a final state $t$ will be reached?  The reachability question
using the \texttt{reach} predicate is defined in
Figure~\ref{fig:runex}(b).  Consider the likelihood of reaching state
\texttt{s3} from \texttt{s0}.  This query can be posed as the
predicate \texttt{prob(reach(s0, 0, s3), P)}, where \texttt{prob/2}
finds the probability of answers (\texttt{P}) to a given query
\texttt{reach(s0,0,s3)}.

%The language $\mathtt{s0}^+\mathtt{s1}*s3$ gives the set of all
%executions that start at \texttt{s0} and end at \texttt{s3}.  

The query \texttt{prob(reach(s0,0,s3),P)} cannot be evaluated in
PRISM.  We illustrate this by first describing PRISM's inference at a
high level.  In PRISM, inference of probabilities proceeds in the same
way as logical inference, except when the selected literal is an
\texttt{msw}.  In this case, the inference procedure enumerates the
values of the random variable, and continues the inference for each
value (by backtracking).  The probability of a derivation is simply
the product of the probabilities of the random variables (\texttt{msw}
outcomes) used in that derivation (under the independence assumption).
The probability of a query answer is the
sum of probabilities of the set of all derivations for that answer
(using the mutual-exclusiveness and finiteness assumptions).
Note that \texttt{reach(s0,0,s3)} has infinitely many derivations, and
hence PRISM cannot infer its probability.

Markov chains can be encoded in ProbLog and LPAD~\cite{lpad} in a similar manner.
As is the case for PRISM, however, analogous reachability queries
cannot be evaluated in these systems either.  The sequence of
random-variable valuations used in the derivation of an answer is
called an \emph{explanation}.  In contrast to PRISM,
ProbLog~\cite{Problog} and PITA~\cite{RiguzziSwift10a}, which is an
implementation of LPAD, materialize the set of explanations of an
answer in the form of a BDD.  Probabilities
are subsequently computed based on the BDD.  This approach permits
these systems to correctly infer probabilities even when the
independence and mutual-exclusion assumptions are violated.  Note that
in the evaluation of \texttt{reach}, when a state is encountered, the
next state is determined by a fresh random process.  Hence, the
\emph{set of explanations} of \texttt{reach(s0,0,s3)} is
infinite.\footnote{This is in contrast to the link-analysis examples
  used in ProbLog and PITA~\cite{RiguzziSwift10b}, where, even though
  the number of derivations for an answer may be infinite, the number of
  explanations is finite.}
Since BDDs can only represent finite sets, the probability of
\texttt{reach(s0,0,s3)} cannot be computed in ProbLog or LPAD.

To correctly infer the probability of \texttt{reach(s0,0,s3)}, we
need an algorithm that works even when the set of explanations is
infinite.  Moreover, it is easy to construct queries where the
independence and mutual exclusion properties do not hold.  
For example, consider the problem of
inferring the probability of reaching \texttt{s3} or \texttt{s4}
(i.e., the query \texttt{reach(s0,0,s3); reach(s0,0,s4)}).  Since some
paths to \texttt{s3} pass through \texttt{s4}, explanations for
\texttt{reach(s0,0,s3)} and \texttt{reach(s0,0,s4)} are not mutually
exclusive.  The example of Fig.~\ref{fig:runex} illustrates that to
build model checkers based on PLP, we need an inference
algorithm that works even when the finiteness, mutual-exclusion and
independence assumptions are simultaneously violated.

\newcommand{\newalg}{PIP\xspace}

\paragraph{Summary of Contributions:}  In this paper, we present
\newalg (for ``Probabilistic Inference Plus''), a new algorithm for
inferring probabilities of queries in a probabilistic logic program.
\newalg is applicable even when
explanations are not necessarily mutually exclusive or independent,
and the number of explanations is infinite.  We demonstrate the
utility of this new inference algorithm by constructing model
checkers for a rich class of probabilistic models and temporal logics
(see Section~\ref{sec:applications}).
\comment{
including those for PCTL properties of Discrete-Time Markov chains;
reachability properties in Recursive Markov Chains; and GPL properties
of Reactive Probabilistic Transition Systems.}
Our model checkers are
based on high-level, logical encodings of the semantics of the process
languages and temporal logics, thus retaining the highly declarative
nature of our prior work on model checking non-probabilistic
systems.

We have implemented our \newalg inference algorithm in XSB
Prolog~\cite{XSB}.  
\comment{Based on this implementation, we have encoded
probabilistic model checkers for a variety of temporal logics, including
PCTL$^*$ and GPL (which subsumes PCTL$^*$), and for process languages of
varying complexity (from Reactive Modules to Recursive Markov Chains).
}
Our experimental results show that, despite the highly declarative nature
of our encodings of the model checkers, their performance is 
competitive with their native implementations.

The rest of this paper develops along the following lines.
Section~\ref{sec:prelim} provides requisite background on probabilistic
logic programming.  Section~\ref{sec:resolve} presents our \newalg
algorithm.
%, our new algorithm for extended probabilistic inference.
Section~\ref{sec:applications} describes our PLP encodings of
probabilistic model checkers, while Section~\ref{sec:results}
contains our experimental evaluation.  Section~\ref{sec:conc}
offers our concluding remarks and directions for future work.

\section{Related Work}
\label{sec:related}

%Works on LP-based model checking

There is a substantial body of prior work on encoding complex model
checkers as logic programs.  These approaches range from using
constraint handling to represent sets of states such as those that arise
in timed systems~\cite{PonGup97,DRS:RTSS00,MP:CL00,PRR:ICLP02}, 
data-independent systems~\cite{SR:ICFEM03} and other infinite-state
systems~\cite{DelzannoPodelski:99,MP:FSTTCS99};
tabling to handle fixed point computation~\cite{XMC:CAV00,Farwer-Leuschel:04};
procedural aspects of proof search to handle name handling~\cite{YRS:STTT04}
and greatest fixed points~\cite{Gupta-coinduction:07}.  However, all these
works deal only with non-probabilistic systems.

\comment{ 
Moreover, most of these works exploited existing logic programming techniques to implement model checkers for novel systems.  In contrast, we find that existing techniques for probabilistic inference in PLP is insufficient for the model checking of probabilistic systems.  This paper presents PIP which is applicable to PLPs in general, and also enables model checkers for probabilistic systems to be constructed at the same high level as those for non-probabilistic systems.
}

% Works on probabilistic logic programming

With regard to related work on probabilistic inference,
Statistical Relational Learning (SRL) has emerged as a rich area of research
into languages and techniques for supporting modeling, inference and learning
using a combination of logical and statistical methods~\cite{srlbook}.
Some SRL techniques, including Bayesian Logic Programs (BLPs)~\cite{blp},
Probabilistic Relational Models (PRMs)~\cite{prm} and Markov Logic Networks
(MLNs)~\cite{mln}, use logic to compactly represent statistical models.
Others, such as PRISM~\cite{sato-kameya-prism97}, Stochastic Logic Programs
(SLP)~\cite{Muggleton}, Independent Choice Logic (ICL)~\cite{PooleICL},
CLP(BN)~\cite{clpbn}, ProbLog~\cite{Problog}, LPAD~\cite{lpad} and
CP-Logic~\cite{CPlogic}, define inference primarily in logical terms,
subsequently assigning statistical properties to the proofs.  
Motivated primarily by knowledge representation problems, these works 
have been naturally restricted to cases
where the models and the inference proofs are finite.  Recently, a number of
techniques have
generalized these frameworks to handle
random variables that range over continuous
domains (e.g.~\cite{hblp,hprm,hmln,hproblog,apprProblog,contdist-inference}), but still restrict proof structures to be finite.

% Works on probabilistic model checking

Modeling and analysis of probabilistic systems, both discrete- and
continuous-time, has been an actively researched area.  Probabilistic
Computation Tree Logic (PCTL)~\cite{pCTL} is a widely used temporal logic for
specifying properties of discrete-time probabilistic systems.
PCTL$^*$~\cite{PCTLstar} is a probabilistic extension of LTL and is more
expressive than PCTL.  Generalized Probabilistic Logic (GPL)~\cite{GPL} is a
probabilistic variant of the modal mu-calculus.  The Prism model
checker~\cite{PRISMMC} is a leading tool for modeling and verifying a wide
variety of probabilistic systems: Discrete- and Continuous-Time Markov chains
and Markov Decision Processes.  There is also prior work on techniques for
verifying more expressive probabilistic systems, including Recursive Markov
chains (RMCs)~\cite{RMC} and Probabilistic Push-Down systems~\cite{PPDS}, both
of which exhibit context-free behavior.  The probability of reachability properties in such systems is computed as the least solution to a corresponding set of monotone polynomial equations.  PReMo~\cite{PREMO} is a model checker for RMCs.  Reactive Probabilistic Labeled Transition Systems (RPLTS)~\cite{GPL} generalize Markov chains by adding external choice (multiple labeled actions).  GPL properties of such systems are also computed as the least (or greatest, based on the property) solution to a set of monotone polynomial equations.  To the best of our knowledge, this paper presents the first implementation of a GPL model checker.

\section{Preliminaries}
\label{sec:prelim}

\paragraph{Notations:}
The root symbol of a term $t$ is denoted by $\troot(t)$ and its $i$-th
subterm by $\arg_i(t)$.  Following traditional LP notation, a term
with a predicate symbol as root is called an \emph{atom}.  The set of
variables in a term $t$ is denoted by $\vars(t)$.  A term $t$ is
\emph{ground} if $\vars(t)=\emptyset$. 

Following PRISM, a \emph{probabilistic logic program} (PLP) is of the
form $P = P_F \cup P_R$, where $P_R$ is a definite logic program, and
$P_F$ is the set of all possible \texttt{msw/3} atoms.  The set of
possible \texttt{msw} atoms and the distribution of their subsets is given
by \texttt{values} and \texttt{set\_sw} directives, respectively.  For
example, clauses \texttt{trans} and \texttt{reach} in
Fig.~\ref{fig:runex}(b) are in $P_R$.  The set $P_F$ of that program
contains \texttt{msw} atoms such as \texttt{msw(t(s0), 0, s0)},
\texttt{msw(t(s0),next(0), s0)}, \texttt{msw(t(s0), 0, s1)},
$\ldots$, \texttt{msw(t(s1),next(0), s1)}, $\ldots$.

In an atom of
the form $\kw{msw}(t_1, t_2, t_3)$, $t_1$ is a term representing a
random process (switch in PRISM terminology), $t_2$ is an instance and
$t_3$ is the outcome of the process at that instance.  According to
PRISM semantics, two \texttt{msw} atoms with distinct processes are
independent; and two \texttt{msw} atoms with distinct instances (even
if they have the same process) are independent.  Two \texttt{msw}
atoms with the same process and instance but different outcomes are
mutually exclusive.

\newcommand{\expl}{{\cal E}}
\section{The Inference Procedure PIP}
\label{sec:resolve}

\newcommand{\temporal}[1]{{\mathtt{temporal}(#1)}}
\newcommand{\predicates}[1]{{\mathtt{preds}(#1)}}

A key idea behind the \newalg inference algorithm is to represent the
(possibly infinite) set of explanations in a symbolic form.  Observe
from the example in Fig.~\ref{fig:runex} that, even though the set of
paths (each with its own distinct probability) from state
\texttt{s0} to state \texttt{s3} is infinite, the regular expression
$\kw{s0}^+\kw{s1}^*\kw{s4}^?\kw{s3}$ captures this set
exactly.   Following this analogy, we devise a grammar-based
notation that can succinctly represent infinite sets of finite
sequences.

\begin{Def}[Explanation]
An \emph{explanation} of an atom $A$ with respect to a PLP
$P = P_F \cup P_R$ is a set $\xi \subseteq P_F$ of \kw{msw}
atoms such that (i)~$\xi, P_R \vdash A$ and (ii)~$\xi$ is consistent,
i.e.\ it contains no pair of mutually exclusive \kw{msw} atoms.

The set of all explanations of $A$ w.r.t.\ $P$ is denoted by
$\expl_{P}(A)$.
\hfill $\Box$
\label{def:explanation1}
\end{Def}

\comment{
Before we can construct a data structure to succinctly represent
explanations, we need to consider the 
use of ``instance identifiers'' (e.g., the \texttt{I} in
\texttt{msw(t(s0), I, s1)}).
}

\begin{Ex}[Set of explanations]
Consider the PLP of Fig.~\ref{fig:runex}(b).  The set of explanations for
\texttt{reach(s0,0,s3)} is:
\[
\begin{array}[]{l}
\kw{msw}(t(s0), 0, s1),\ \kw{msw}(t(s1), \id{next}(0), s3).\\  
\kw{msw}(t(s0), 0, s0),\ \kw{msw}(t(s0), \id{next}(0), s1),\ \kw{msw}(t(s1),
  \id{next}(\id{next}(0)), s3).\\  
\vdots\\
\kw{msw}(t(s0), 0, s1),\ \kw{msw}(t(s1), \id{next}(0), s1),\ \kw{msw}(t(s1),
  \id{next}(\id{next}(0)), s3).\\  
\vdots
\end{array}
\]
\label{ex:set-of-explanations}
\end{Ex}
\vspace*{-3em}

\subsection{Representing Explanations}
As Example~\ref{ex:set-of-explanations} illustrates, a representation
in which instance identifiers are explicitly captured will not be
nearly as compact as the corresponding regular expression (shown
earlier).  On the other hand, a representation (like the regular
expression) that completely ignores instance identifiers will not be
able to identify identical instances of a random process nor properly
distinguish distinct ones.

We solve this problem by observing that in PRISM's semantics,
different instances of the same random process are \emph{independent
  and identically distributed} (i.i.d.).  Consequently, the
probability of \texttt{reach(s0,0,s3)} (reaching \texttt{s3} from
\texttt{s0} starting at instance \texttt{0}), is the same as that of
\texttt{reach(s0,next(0),s3)} (starting at instance \texttt{1}), which
is the same as that of \texttt{reach(s0,H,s3)}, for any instance
\texttt{H}.  Consequently, it is sufficient to infer probabilities for
a single parameterized instance.
Below, we formalize the set of PLP programs for which such an
abstraction is possible.

\begin{Def}[Temporal PLP]
A \emph{temporal} probabilistic logic program is a probabilistic logic
program $P$ with declarations of the form
\texttt{temporal($p/n-i$)}, where $p/n$ is an $n$-ary predicate, and
$i$ is an argument position (between 1 and $n$) called the
\emph{instance argument} of $p/n$ . Predicates $p/n$ in such declarations
are called \emph{temporal predicates}.
\hfill $\Box$
\label{def:temporalPLP}
\end{Def}

The set of temporal predicates in a temporal PLP $P$ is denoted
by $\temporal{P}$; the set of all predicates in $P$ is denoted by
$\predicates{P}$.  
By convention, every temporal PLP
contains an implicit declaration \texttt{temporal(msw/3-2)},
indicating that \texttt{msw/3} is a temporal predicate, and its
second argument is its instance argument.  The instance argument of a
predicate $p/n$ is denoted by $\inst(p/n)$.  
For example, the program of Fig.~\ref{fig:runex}(b) becomes a temporal
PLP when 
\texttt{temporal(trans/3-2)} and \texttt{temporal(reach/3-2)}
are added.  For
this program, 
$\temporal{P} = \{\mathtt{reach/3}, \mathtt{trans/3}, \mathtt{msw/3} \}$,
and $\inst(\mathtt{reach/3}) = \inst(\mathtt{trans/3}) =
\inst(\mathtt{msw/3}) = 2$.

%\noindent
We extend the notion of instance argument from predicates to atoms as
follows.  Let $\alpha$ be an atom in a temporal PLP such that
its root symbol is a temporal predicate, i.e., $\troot(\alpha) \in
\temporal{P}$.  Then the \emph{instance of $\alpha$}, denoted by
$\inst(\alpha)$ by overloading the symbol $\inst$, is
$\arg_{\inst(\troot(\alpha))}(\alpha)$.
We also denote, by $\noninst(\alpha)$, a term constructed by
\emph{omitting} the instance of $\alpha$; i.e.
if $\alpha = f(t_1, \ldots, t_{i-1}, t_i, t_{i+1}, \ldots t_n)$
and $\inst(\alpha) = t_i$, then 
$\noninst(\alpha) = f(t_1, \ldots, t_{i-1}, t_{i+1}, \ldots t_n)$.

Explanations of a temporal PLP can be represented by a notation
similar to Definite Clause Grammars (DCGs).  

\begin{Ex}[Set of explanations using DCG notation]
Considering again the program of Fig.~\ref{fig:runex}(b),
the set of explanations for \texttt{reach(s0,H,s3)}
can be succinctly represented by the following DCG:
\[
\begin{array}[]{l}
\kw{expl}(\kw{reach}(s0, s3),H) \longrightarrow [\kw{msw}(t(s0), H, s0)],
\kw{expl}(\kw{reach}(s0,s3),next(H)).\\
\kw{expl}(\kw{reach}(s0, s3),H) \longrightarrow [\kw{msw}(t(s0), H, s1)],
\kw{expl}(\kw{reach}(s1,s3),next(H)).\\
\kw{expl}(\kw{reach}(s1, s3),H) \longrightarrow [\kw{msw}(t(s1), H, s1)],
\kw{expl}(\kw{reach}(s1,s3),next(H)).\\
\kw{expl}(\kw{reach}(s1, s3),H) \longrightarrow [\kw{msw}(t(s1), H, s3)],
\kw{expl}(\kw{reach}(s1,s3),next(H)).\\
\kw{expl}(\kw{reach}(s1, s3),H) \longrightarrow [\kw{msw}(t(s1), H, s4)],
\kw{expl}(\kw{reach}(s4,s3),next(H)).\\
\kw{expl}(\kw{reach}(s3,s3),H) \longrightarrow [].\\
\kw{expl}(\kw{reach}(s4, s3),H) \longrightarrow [\kw{msw}(t(s4), H, s3)],
\kw{expl}(\kw{reach}(s3,s3),next(H)).\\
\end{array}
\]
\label{ex:explanations-with-dcg}
\end{Ex}

Note that each \texttt{expl} generates a sequence of \texttt{msw}s.
For this example, it is also the case that in a string generated from
$\kw{expl}(\kw{reach}(s0,s3), H)$, the \texttt{msw}s all have instances
equal to or later than \texttt{H}.  It is then immediate that
\texttt{msw(t(s0), H, s0)} is independent of \emph{any} \texttt{msw}
generated from $\kw{expl}(\kw{reach}(s0, s3), next(H))$.  This property
holds for an important subclass called \emph{temporally well-formed
programs}, defined as follows.

%Definition of PLPs, following PRISM

\begin{Def}[Temporally Well-Formed PLP]
  A temporal PLP $P$ is said to be \emph{temporally well
  formed} if for each clause $(\alpha \when \beta_1, \ldots \beta_n) \in
    P$:
    \begin{itemize}
    \item If $\troot(\alpha) \in \temporal{P}$ then
      $\forall\ i, 1 \leq i \leq n,
      \mbox{ s.t. } \troot(\beta_i) \in \temporal{P}$, $\inst(\beta_i)$
      contains $\inst(\alpha)$, and $\vars(\beta_i) = \vars(\alpha)$.
    \item If $\troot(\alpha) \not \in \temporal{P}$ then there is at
      most one $i$, $1 \leq i \leq n$ s.t.\ $\troot(\beta_i) \in \temporal{P}$.
    \item Instance arguments $\inst(\alpha)$ or
    $\inst(\beta_i)$ or their subterms are unified only with other
    instance arguments, their subterms, or with ground terms. 
\hfill $\Box$
    \end{itemize}
\label{def:temporal-wf-PLP}
\end{Def}

For temporally well-formed programs, the explanations for an atom can
be represented succinctly by DCGs.  Such DCGs 
are called explanation generators.
%defined below.

\begin{Def}[Explanation Generator]
Let $P$ be a temporally well-formed PLP and let $Q$ be a query such
that all non-instance-arguments of $Q$ are ground (i.e.\ $\noninst(Q)$
is ground).
Then, an \emph{explanation generator} for $Q$ with respect to $P$ is a
DCG $\Gamma$, with non-terminals of the form
$\kw{expl}(G,H)$ and terminals of the form $\kw{msw}(r,H,v)$ such that:
\begin{itemize}
\item For every production $(\beta_0 \rightarrow \beta_1, \ldots, \beta_n) \in
  \Gamma$, $\forall\ i$, $0\leq i\leq n$, all non-instance-arguments of
  $\beta_i$ are ground; i.e., if $\beta_i = \kw{expl}(G,H)$, then
  $G$ is ground, and if $\beta_i = \kw{msw}(r,H,v)$, then $r$ and $v$
  are ground.
\item $\expl_{P}(Q)$, the set of explanations for $Q$ w.r.t.\ $P$, 
  is identical to the language of $\Gamma$ with
  $\kw{expl}(\noninst(Q), \inst(Q))$
  as the start symbol.
\hfill $\Box$
\end{itemize}
\label{def:explanation-generator}
\end{Def}

%\begin{Ex}
  The DCG in Example~\ref{ex:explanations-with-dcg} is the explanation
  generator for the query \texttt{reach(s0,H,s3)} over the program given
  in Figure~\ref{fig:runex}(b). 
%\end{Ex}
%
In general, an explanation generator may not be in a form from which
we can directly infer the probabilities.  For this purpose, we use the
\emph{factoring} algorithm described below.

\subsection{Factored Explanation Diagrams}

The factored form of an explanation generator is obtained by
constructing a Factored Explanation Diagram (FED), whose structure closely
follows that of a BDD.  Similar to a BDD, a FED is a labeled direct
acyclic graph with two distinguished leaf nodes: \kw{tt}, representing
\id{true}, and \kw{ff}, representing \id{false}.  While the internal
nodes of a BDD are Boolean variables, a FED contains two kinds of
internal nodes: one representing terminal symbols of
explanations (\kw{msw}s), and the other representing non-terminal
symbols of explanations (\kw{expl}s).  We use a 
partial order among nodes, denoted by ``$<$'', to construct a FED.

\begin{Def}[Factored Explanation Diagram]
  A \emph{factored explanation diagram} (FED) is a labeled directed
  acyclic graph with:
  \begin{itemize}
  \item Four kinds of nodes: \kw{tt}, \kw{ff}, $\kw{msw}(r, h)$ and
    $\kw{expl}(t,h)$, where $r$ is a ground term representing a random
    process, $t$ is a ground term, and $h$ is an instance term;
 \item Nodes \kw{tt} and \kw{ff} are 0-ary, and occur only at leaves
    of the graph;
  \item $\kw{msw}(r,h)$ is an $n$-ary node when $r$ is random process
    with $n$ outcomes, and the edges to the $n$ children are labeled
    with the possible outcomes of $r$;
  \item $\kw{expl}(t,h)$ is a binary node, and the edges to the
    children are labeled $0$ and $1$.
 \item If there is an edge from node $x_1$ to $x_2$, then $x_1 <
   x_2$.\hfill $\Box$
 \end{itemize}
  \label{def:fed}
\end{Def}
Note that the multi-valued decision diagrams used in the implementation
of PITA~\cite{RiguzziSwift10b} are a special case of FEDs with only \kw{tt},
\kw{ff} and $\kw{msw}(r,h)$ nodes, where $r$ and $h$ are ground.

\sloppypar
We represent non-trivial FEDs by $x?\id{Alts}$,
where $x$ is the node and $\id{Alts}$ is the list of edge-label/child pairs. 
For example, a FED $F$ whose root is an MSW node
is written as $\kw{msw}(r,h)?[\alt{v_1}{F_1}, \alt{v_2}{F_2}, \ldots, \alt{v_n}{F_n}]$, where
$F_1, F_2, \ldots, F_n$ are children FEDs (not all necessarily distinct)
and $v_1, v_2, \ldots, v_n$ are possible outcomes of the random process $r$ 
such that $v_i$ is the label on the edge from $F$ to $F_i$.
Similarly, a FED $F$ whose root is an EXPL node is written as 
$\kw{expl}(t, h)?[\alt{0}{F_0}, \alt{1}{F_1}]$, where $F_0$ and $F_1$ are the
children of $F$ with edge labels $0$ and $1$, respectively.

We now define the ordering relation ``$<$'' among nodes. 
We first define a time order ``$\prec$'' among instances such that
$h_1 \prec h_2$ if $h_1$ represents an earlier time instant than
$h_2$.  If $h_1 \not \preceq h_2$ and $h_2 \not \preceq h_1$, then
$h_1$ and $h_2$ are incomparable, denoted as $h_1 \sim h_2$.
We also assume an arbitrary order $<$ among terms.
\begin{Def}[Node order]
  Let $x_1$ and $x_2$ be nodes in a FED.  Then $x_1 < x_2$ if it
  matches one of the following cases:
  \begin{itemize}
  \item $\kw{msw}(r_1, h_1) < \kw{msw}(r_2, h_2) $ if $h_1 \prec h_2$
or ($r_1 < r_2$ and  ($h_1 = h_2$ or $h_1 \sim h_2$))
 \item $\kw{msw}(r_1, h_1) < \kw{expl}(t_2, h_2)$ if $h_1 \prec h_2$
   or $h_1 \sim h_2$
 \item $\kw{expl}(t_1, h_1) < \kw{expl}(t_2, h_2)$ if $t_1 < t_2$ and
   $h_1 \sim h_2$. \hfill $\Box$
  \end{itemize}
\label{def:node-order}
\end{Def}

\vspace*{-1em}
\begin{Pro}[Independence and Node Order]
  For any nodes $x_1$, $x_2$ in an FED, if $x_1 < x_2$ or $x_2 < x_1$,
  then $x_1$ and $x_2$ are independent.
\label{pro:independence}
\end{Pro}

\comment{
\subsubsection{Node Ordering in FEDs}

We now consider the partial order among nodes that is used in the
construction of FEDs.  The key idea is to ensure that for all nodes
$x_1$ and $x_2$, whenever $x_1 < x_2$ or $x_2 < x_1$, $x_1$ and $x_2$ are
independent.  We consider a time order ``$\prec$'' among instances,
and standard term order (e.g., Prolog's \verb+@<+)  among terms to
construct our node order.  We say that $h_1 \prec h_2$ if $h_1$
represents an earlier time instant than $h_2$.   If $h_1 \not \preceq
h_2$ and $h_2 \not \preceq h_1$, we say that $h_1$ and $h_2$ are
incomparable, and denote this as $h_1 \sim h_2$.
We say that $\kw{msw}(r_1,h_1) < \kw{msw}(r_2,h_2)$ if
%and only if
either (a)~$h_1 \prec h_2$ or (b)~$r_1 < r_2$ (where $<$ is the
term order) and $h_1=h_2$ or $h_1 \sim h_2$.   Note that if $x_1 =
\kw{msw}(r_1,h_1)$ and $x_2 =  \kw{msw}(r_2,h_2)$, then $x_1 < x_2$,
$x_2 < x_1$, or $x_1 = x_2$; i.e., ``$<$'' defines a total order
among \texttt{msw} nodes, analogous to the total order among variables
in a BDD. 

The ordering between \kw{expl} and \kw{msw}
nodes is more subtle.   Note that
$\kw{expl}(t,h)$ may generate a random process from instance $h$ or
later.  Hence, if we compare two nodes $x_1 = \kw{msw}(r,h_1)$ and
$x_2 = \kw{expl}(t_2, h_2)$ such that $h_1 \prec h_2$, then the
explanations of $\kw{expl}(t_2, h_2)$ will be independent of the
valuation of random process $r$ at $h_1$, and we order $x_1 < x_2$.

On the other hand, if we compare two nodes $x'_1 = \kw{msw}(r,h'_1)$ and
$x'_2 = \kw{expl}(t_2,  h'_2)$ such that $h'_2 \prec h'_1$, then it is
possible that the derived process $t_2$ uses random process $r$ at
time $h'_1$, and hence the two nodes $x'_1$ and $x'_2$ are not
independent!  In this case, we can say neither that $x'_1 < x'_2$, nor
$x'_2 < x'_1$.   Finally, if $h'_1 \sim h'_2$,
we can order the
two nodes $x'_1$ and $x'_2$ in any way.  In such cases, we find it
convenient to  order the \kw{msw} node $x'_1$ to be $< x'_2$, the
\kw{expl} node.

A similar problem arises with the ordering between two \kw{expl} nodes.
If we compare two nodes $x_1 = \kw{expl}(t_1,h_1)$ and $x_2 =
\kw{expl}(t_2, h_2)$ such that $h_1 \prec h_2$, it is still possible
for the derived process $t_1$ to use the same basic random processes
at the same instance as process $t_2$.  Hence we can order two
\kw{expl} nodes only if their instances describe distinct branches in
time ($h_1 \sim h_2$).  We then use the term order between $t_1$ and
$t_2$, and an arbitrary lexicographic order between $h_1$ and $h_2$ to order
$x_1$ and $x_2$.
}

%% \subsubsection{Operations on FEDs}
%% 
%% Conjunction and disjunction of two FEDs is defined as follows.

\vspace*{-1em}
\begin{Def}[Binary Operations on FEDs]
$F_1 \oplus F_2$, where $\oplus \in \{\land, \lor\}$ is a FED $F$
derived as follows:
\begin{itemize}
\item $F_1$ is \kw{tt}, and $\oplus = \lor$, then $F =
  \kw{tt}$.
\item $F_1$ is \kw{tt}, and $\oplus = \land$, then $F =
  F_2$.
\item $F_1$ is \kw{ff}, and $\oplus = \land$, then $F =
  \kw{ff}$.
\item $F_1$ is \kw{ff}, and $\oplus = \lor$, then $F =
  F_2$.
\item $F_1 = x_1?[\alt{v_{1,1}}{F_{1,1}}, \ldots, \alt{v_{1,n_1}}{F_{1,n_1}}]$, 
$F_2 = x_2?[\alt{v_{2,1}}{F_{2,1}}, \ldots, \alt{v_{2,n_2}}{F_{2,n_2}}]$:
\begin{description}
\item[c1.] $x_1 < x_2$:   $F = x_1?[\alt{v_{1,1}}{(F_{1,1} \oplus F_2)}, \ldots,
  \alt{v_{1,n_1}}{(F_{1,n_1} \oplus F_2)}]$ 
\item[c2.] $x_1 = x_2$:   $F = x_1?[\alt{v_{1,1}}{(F_{1,1} \oplus F_{2,1})}, \ldots,
  \alt{v_{1,n_1}}{(F_{1,n_1} \oplus F_{2,n_1})}]$ 
\item[c3.] $x_1 > x_2$:   $F = x_2?[\alt{v_{2,1}}{(F_1 \oplus F_{2,1})}, \ldots,
  \alt{v_{2,n_2}}{(F_1 \oplus F_{2,n_2})}]$ 
\item[c4.] $x_1 \not<x_2, x_2 \not< x_1$: $F = \kw{expl}(\kw{merge}(\oplus,
  F_1, F_2), h)?[\alt{0}{\kw{ff}}, \alt{1}{\kw{tt}}]\quad$ where $h$
  is the common part of instances of $x_1$ and $x_2$.\ \hfill $\Box$
\end{description}
\end{itemize}
\label{def:fed-ops}
\end{Def}
Note that Def.~\ref{def:fed-ops} is a generalization of the corresponding
operations on BDDs.  Also, when $x_1$ and $x_2$ are both
\texttt{msw} nodes, since $<$ defines a total order between them, case
\textbf{c4} will not apply.  
When the operand nodes cannot be ordered (case \textbf{c4}), we
generate a placeholder (a \texttt{merge} node) indicating the operation
to be performed.  Such placeholders will be expanded when FEDs are built from
an explanation generator (see Def.~\ref{def:fed-construction} below).
Note that \texttt{merge} nodes may be generated
only if one or both arguments is a FED rooted at an \texttt{expl} node.

We now give a procedure for constructing FEDs from an
explanation generator for query $Q$ with respect to program $P$.

\begin{Def}[Construction of Factored Explanation Diagrams]
%% CR -- something wrong here with $P$
Given an explanation generator $\Gamma$, $F$ is a FED
corresponding to goal $G$  if $\kw{fed}(G, F)$ holds,
where \kw{fed\_c} is the smallest relation and
$E$ is the smallest set such that:
\begin{itemize}
\item $\kw{fed\_c}(\beta_0, F)$ holds whenever 
  $\{(\beta_0 \rightarrow \beta_{1,1},\ldots, \beta_{1,n_1}),$
  $(\beta_0 \rightarrow \beta_{2,1},\ldots, \beta_{2,n_2}),$ $ \ldots,$
  $(\beta_0 \rightarrow \beta_{k,1},\ldots, \beta_{k,n_k})\}$ is the set of
  all clauses in $P$ with $\beta_0$ on the left hand side, and
\[
F = \bigvee_{i=1}^{k} \quad \bigwedge_{j=1}^{n_k}\quad F_{i,j} \quad\quad
\mbox{where}\ F_{i,j} \ \mbox{is such that}\ \kw{fed}(\beta_{i,j}, F_{i,j})\ \mbox{holds}
\]

\item $\kw{fed\_c}(\beta_0, F)$ holds whenever
  $\beta_0 = \kw{merge}(\oplus, F_1, F_2) \in E$, and

  \begin{itemize}
  \item   $F_1 = \kw{expl}(t_1, h_1)?[\alt{0}{F_{1,0}}, \alt{1}{F_{1,1}}]$,
    \quad $\kw{fed}(\kw{expl}(t_1,h_1), F_1')$ holds,\\ and
    $F = \oplus(F_1'[\kw{ff} \mapsto F_{1,0}, \kw{tt} \mapsto F_{1,1}],  F_2)$.
    
  \item   $F_2 = \kw{expl}(t_2, h_2)?[\alt{0}{F_{2,0}}, \alt{1}{F_{2,1}}]$,
    \quad $\kw{fed}(\kw{expl}(t_2,h_2), F_2')$ holds,\\ and
    $F = \oplus(F_1, F_2'[\kw{ff} \mapsto F_{2,0}, \kw{tt} \mapsto F_{2,1}])$.    
 \end{itemize}

\item $\kw{fed}(G, F)$ holds whenever
  \begin{itemize}
  \item $G=\kw{msw}(r,h,v)$ and $F=\kw{msw}(r,h)?[\alt{v_1}{F_1}, \ldots
    \alt{v_n}{F_n}]$ where for all $i$, 
    $F_i = \kw{tt}$ if $v_i = v$ and $F_i = \kw{ff}$ otherwise.
  \item $G=\kw{expl}(t,h)$, $h$ is neither a ground term nor
    a variable, \\
    and $F = \kw{expl}(t,h)?[\alt{0}{\kw{ff}},\alt{1}{\kw{tt}}]$.
  \item $G=\kw{expl}(t,h)$, $h$ is either ground or a variable,
    and $F = \lor_{\kw{fed\_c}(G, F')}  F'$.
  \end{itemize}
\item $\kw{merge}(\oplus, F_1, F_2) \in E$ whenever there is some $G, F$
  such that $\kw{fed}(G, F)$ holds, and there is a node in $F$ of the
  form $\kw{expl}(\kw{merge}(\oplus, F_1, F_2), h)$.\ \hfill $\Box$
\end{itemize}
\label{def:fed-construction}
\end{Def}

The above definition is inductive, and can be turned into a tabled
logic program implementing the construction procedure.
Furthermore, FEDs are maintained using a dictionary
to ensure that the FEDs have a DAG instead of tree structure.

\begin{figure}
  \centering
  \begin{tabular}[c]{cc}
\begin{minipage}[c]{1.5in}
  \begin{tabular}[c]{c}
\includegraphics[scale=0.45]{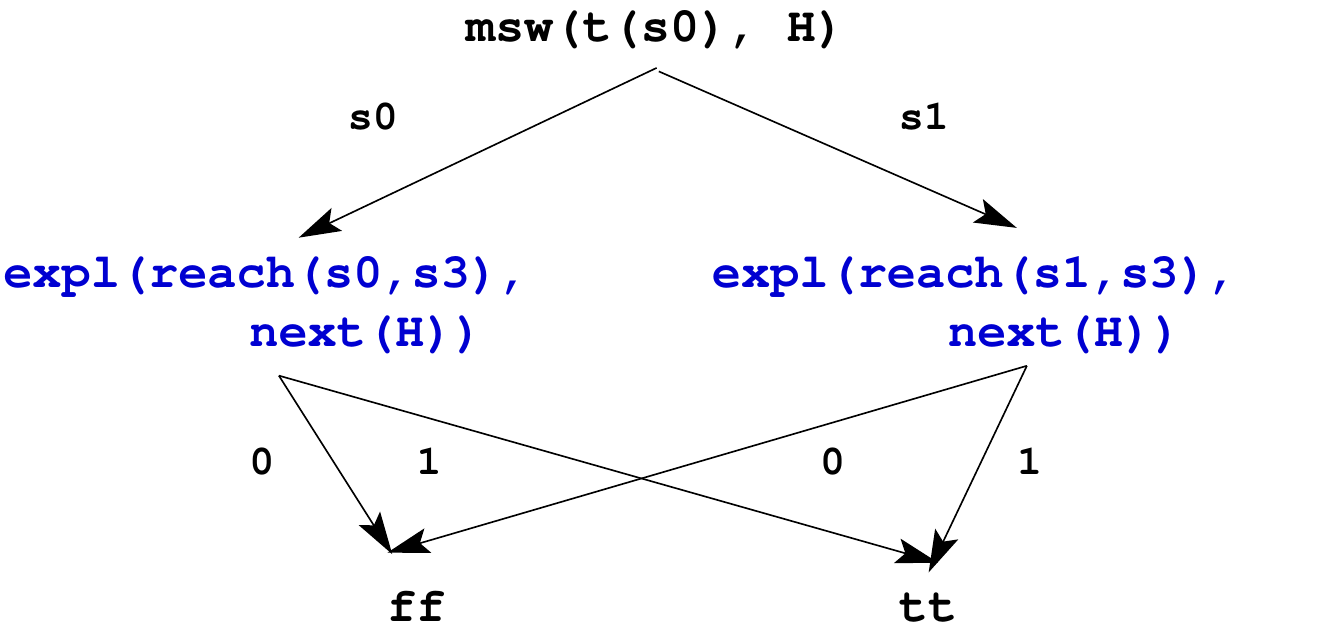}\\
(a) FED for \texttt{expl(reach(s0,s3),H)}\\[0.2in]
\includegraphics[scale=0.45]{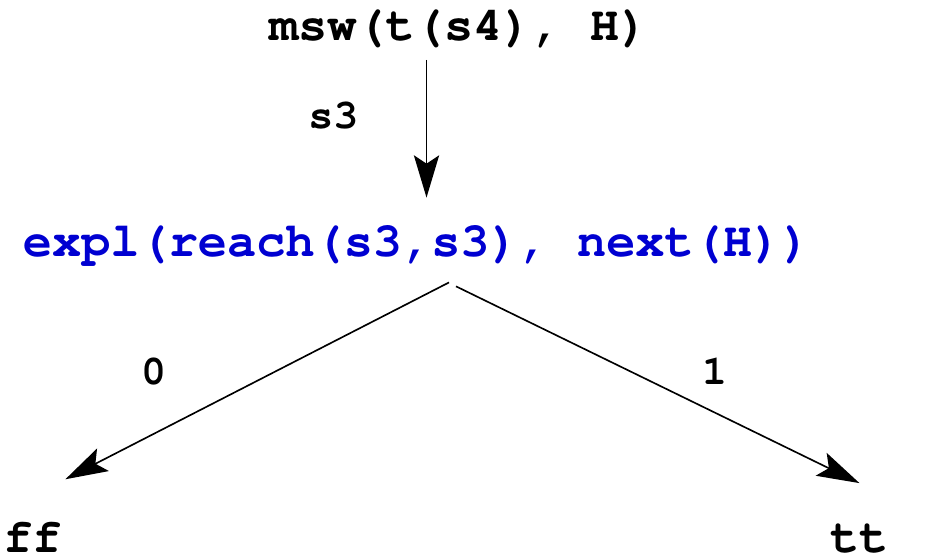}\hspace*{.55in}\\
(c) FED for \texttt{expl(reach(s4,s3),H)}
\end{tabular}
\end{minipage}
& 
\begin{minipage}[c]{3in}
  \begin{tabular}[c]{c}
\\[0.6in]
\includegraphics[scale=0.45]{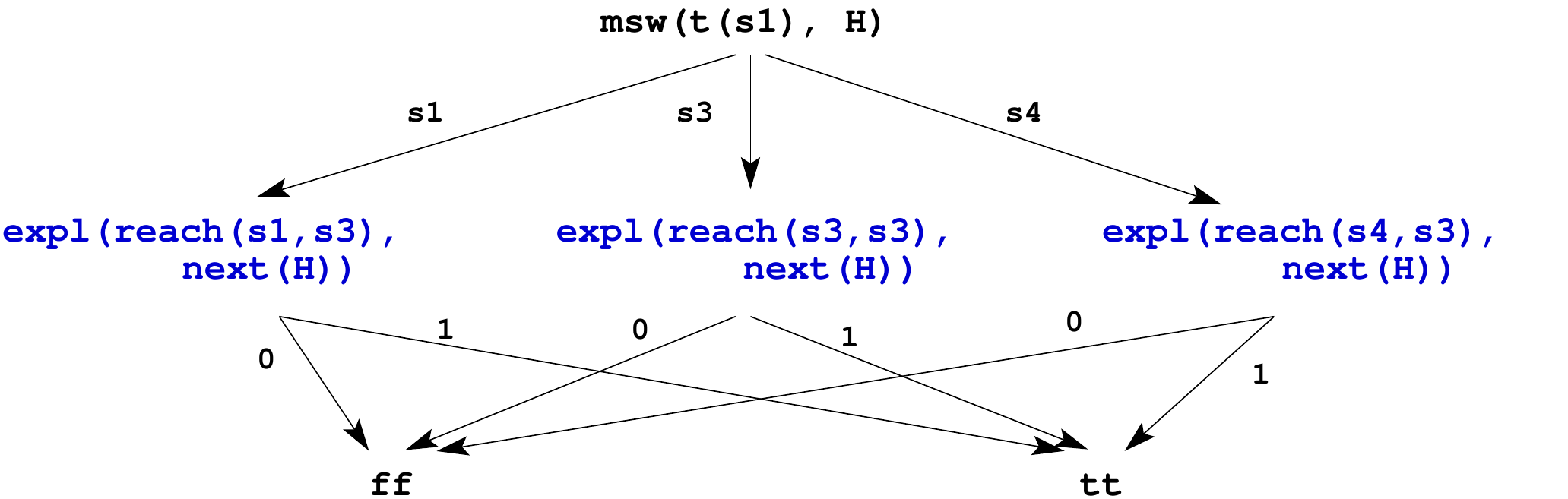}\\
(b) FED for \texttt{expl(reach(s1,s3),H)}
%\includegraphics[scale=0.3]{feds3}\hspace*{1in}\\
%(d)\\
\end{tabular}
\end{minipage}
\end{tabular}
\caption{FEDs for Example~\protect{\ref{ex:explanations-with-dcg}}}
\label{fig:feds}
\end{figure}

\begin{Ex}
  Three of the four FEDs for the explanation generator in 
  Example~\ref{ex:explanations-with-dcg} are
  shown in Fig.~\ref{fig:feds}.  The FED for
  \texttt{expl(reach(s3,s3),H)}, not shown in the figure, is
  \texttt{tt}.
\label{ex:feds}
\end{Ex}

\subsection{Computing Probabilities from FEDs}

A factored explanation diagram can be viewed as a \emph{stochastic grammar}. 
Following~\cite{RMC}, we can generate a set of simultaneous equations
from the stochastic grammar, and find the probability of the language from the
least solution of the equations.  The generation of equations from the
factored representation of explanations is formalized below.

\begin{Def}[Temporal Abstraction]
Given a temporal PLP $P$, the temporal abstraction of a term $t$,
denoted by $\abstr(t)$ is $\noninst(t)$ if $\troot(t) \in \temporal{P}$,
     and $\inst(t)$ is non-ground; and $t$ otherwise.
That is, for a term $t$ with a temporal predicate as root, $\abstr(t)$
replaces its instance argument with a special symbol $\bot$ if that
argument is not ground.
\hfill $\Box$
\label{def:temporal-abstraction}
\end{Def}

\comment{
\begin{Def}[Temporal Abstraction]
Given a temporal PLP $P$, the temporal abstraction of a term $t$,
denoted by $\abstr(t)$ is such that 
\begin{align*}
  \abstr(t) = 
  \begin{cases}
   \noninst(t) & \text{if $\troot(t) \in \temporal{P}$,
     and $\inst(t)$ is non-ground}\\
    t & \text{otherwise} 
  \end{cases}
\end{align*}
That is, for a term $t$ with a temporal predicate as root, $\abstr(t)$
replaces its instance argument with a special symbol $\bot$ if that
argument is not ground.
\hfill $\Box$
\label{def:temporal-abstraction}
\end{Def}
}

\begin{Def}[Distribution]
Let $\rho$ be a random process specified in a PLP $P$.  The
set of values produced by $\rho$ is denoted by $\kw{values}_P(\rho)$.  The
distribution of $\rho$, denoted by $\kw{distr}_P(\rho)$, is a function
from the set of all terms over the Herbrand Base of $P$ to $[0,1]$
such that $\sum_{v \in \kw{values}_P(\rho)} \kw{distr}_P(\rho)(v) = 1 $
\hfill $\Box$
\label{def:distribution}
\end{Def}

\begin{Def}[System of Equations for PLP]
  Let $\Gamma$ be an explanation generator, \texttt{fed} be the
  relation defined in Def.~\ref{def:fed-construction}, $V$ be a
  countable set of variables, and $f$ be a one-to-one function from
  terms to $V$.  The system of polynomial equations $E_{(\Gamma,V,f)}$
  $= \{   (f(\abstr(G)) = {\cal P}(F)) \mid \kw{fed(G,F)} \mbox{
    holds} \}$, where ${\cal P}$ is a function that maps FEDs to
  polynomials, is defined as follows:

\[
  \begin{array}{rcl}
{\cal P}(\kw{ff}) &=& 0\\
{\cal P}(\kw{tt}) &=& 1\\
{\cal P}(\kw{msw}(r,h)?[\alt{v_1}{F_1}, \ldots,
\alt{v_n}{F_n}]) &=& \sum_{i=1}^n \kw{distr}(r)(v) * {\cal P}(F_i)\\
{\cal P}(\kw{expl}(t,h)?[\alt{0}{F_0}, \alt{1}{F_1}]) &= &
   f(\abstr(\kw{expl}(t,h))) * {\cal P}(F_1) \\
& & \quad +    (1-f(\abstr(\kw{expl}(t,h)))) * {\cal P}(F_0)\\
 \end{array}
\]
%\hfill $\Box$
\label{def:ff-to-eqn}
\end{Def}

%\begin{Ex}
  The set of equations for Example~\ref{ex:feds} is shown in Fig.~\ref{fig:eqns}.

%generated from the set of FEDs of  Example~\ref{ex:feds}
%(Fig.~\ref{fig:feds}) 
%\label{fig:explanation-equations} 
%\end{Ex}
 
The implementation of the above definition is such that shared FEDs
result in shared variables in the equation system, thereby ensuring
that every FED is evaluated at most once. 
The correspondence between a PLP in factored form and the set of
monotone equations permits us to compute the probability of query
answers in terms of the least solution to the system of equations.

 \begin{figure}
   \centering
   $
   \begin{array}{l@{\extracolsep{5em}}ll}\hline
     x_0 = t_{00} * x_0 + t_{01} * x_1   & t_{00} = .5  & t_{14} = .5\\
     x_1 = t_{11} * x_1 + t_{13} * x_3 + t_{14}*x_4 & t_{01} = .3 & t_{43}=1\\
     x_3 = 1 & t_{11} =.4 &\\
     x_4 = t_{43} * x_3 & t_{13} = .1 &\\\hline
   \end{array}
   $
   \caption{Set of equations generated from the set of FEDs of Example~\protect{\ref{ex:feds}}}
\label{fig:eqns}
\end{figure}

\begin{Thm}[Factored Forms and Probability]
Let $\Gamma$ be an explanation generator for query $Q$ w.r.t.\ program
$P$.  Let $V$ be a set of variables and let $f$ be a one-to-one function
from terms to $V$.  Then, $X$ is the probability of a query answer $Q$
evaluated over $P$, denoted by $\mathtt{prob}(Q, X)$, if $X$ is the value of the
variable $f(\kw{expl}(\noninst(Q), \inst(Q)))$ in the least solution of
the corresponding set of equations, $E_{(\Gamma,V,f)}$.
\end{Thm}

The proof of the above theorem can be obtained by treating the
explanations in $P$ as strings generated by a 
corresponding stochastic CFG.
Such a correspondence is possible since the explanations are
represented in factored form.
The following properties show that the algorithm for finding
probabilities of a query answer is well defined.

\begin{Pro}[Monotonicity]
  If $\Gamma$ is a definite PLP in factored form, $V$ is a set of variables
  and $f$ is a one-to-one function as required by
  Def.~\ref{def:ff-to-eqn}, then the system of
  equations $E_{(\Gamma,V,f)}$ is monotone in $[0,1]$.
\end{Pro}

Monotone systems have the following important property: 
\begin{Pro}[Least Solution~\cite{RMC}]
  Let $E$ be a set of polynomial equations which is monotone in $[0,1]$.  Then 
  $E$ has a least solution in $[0,1]$.  
  Furthermore, a least solution can be computed to within an arbitrary
  approximation bound by an iterative procedure.  
\label{pro:least-solution}
\end{Pro}

Note that FEDs are non-regular since \kw{expl} nodes may have other
\kw{expl} nodes as children, and hence the resulting equations may be
non-linear.  Proposition~\ref{pro:least-solution} establishes that
probability of query answers can be effectively computed even when the
set of equations is non-linear.

%\begin{Ex}
  The probability of the language of explanations in
  Example~\ref{ex:explanations-with-dcg} (via the equations in
  Fig.~\ref{fig:eqns}) is given by
  the value of $x_0$ in the least solution, which is $0.6$.
%\end{Ex}

\section{Applications}
\label{sec:applications}

We now present two model checkers that demonstrate the utility of the
new \newalg inference technique.

%\subsection{Probabilistic CTL (PCTL)}

\paragraph{\textbf{PCTL:}}  The syntax of an illustrative fragment of PCTL is given by:
\[
\begin{array}{rl}
  \id{SF} \Coloneqq& \kw{prop}(A)\ |\ \kw{neg}(\id{SF}) \ |\
  \kw{and}(\id{SF}_1, \id{SF}_2) 
  |\  \kw{pr}(\id{PF}, \kw{gt}, B)\ |\   \kw{pr}(\id{PF}, \kw{geq}, B)\\
  \id{PF} \Coloneqq &  \kw{until}(\id{SF}_1, \id{SF}_2)\ |\ \kw{next}(\id{SF})
\end{array}
\]
Here, $A$ is a proposition and $B$ is a real number in $[0,1]$.
The logic partitions formulae into \emph{state} formulae
(denoted by $\id{SF}$) and \emph{path} formulae (denoted by $\id{PF}$).  
State formulae are given a
non-probabilistic semantics: a state formula is either true or false
at a state.  For example, formula \texttt{prop(a)} is true at state
$s$ if proposition \texttt{a} holds at $s$; a formula
$\kw{and}(\id{SF}_1, \id{SF}_2)$ holds at $s$ if both $\id{SF}_1$ and
$\id{SF}_2$ hold at
$s$.  The formula $\kw{pr}(\id{PF}, \kw{gt}, B)$ holds at a state $s$
if the probability $p$ of the set of all paths on which the 
path formula $\id{PF}$ holds is such that $p > B$ (similarly, $p \geq
B$ for \kw{geq}).   

The formula $\kw{until}(\id{SF}_1, \id{SF}_2)$ holds on a single given
path $s_0, s_1, s_2, \ldots$ if $\id{SF}_2$ holds on state $s_k$ for
some $k \geq 0$, and $\id{SF}_1$ holds for all $s_i$, $0 \leq i < k$.
Full PCTL has a \emph{bounded} \texttt{until} operator, which imposes
a fixed upper bound on $k$; we omit its treatment since it has a
straightforward non-fixed-point semantics.  The \emph{probability} of
a path formula $\id{PF}$ at a state $s$ is the sum of probabilities of
all paths starting at $s$ on which $\id{PF}$ holds.  This semantics is
directly encoded as the probabilistic logic program given in
Fig.~\ref{fig:pctl-mc}.  Observe that the program is temporally well
formed.  Moreover, observe the use of an abstract instance argument
``\texttt{\_}'' the invocation of \texttt{pmodels/3} from
\texttt{pmodels/2}.  This ensures that an explanation generator can be
effectively computed for any query to \texttt{pmodels/2}.

\begin{figure}
\centering
\small
\begin{tabular}{ll}
\% State Formulae & \% Path Formulae\\
\begin{minipage}[t]{2in}
\begin{verbatim}
models(S, prop(A)) :-
    holds(A).
models(S, neg(A)) :- 
    not models(A).
models(S, and(SF1, SF2):-
    models(S, SF1), 
    models(S, SF2).
models(S, pr(PF, gt, B)) :-
    prob(pmodels(S, PF), P),
    P > B.
models(S, pr(PF, geq, X)) :-
    prob(pmodels(S, PF), Y),
    P >= B.
\end{verbatim}
\end{minipage}
&
\begin{minipage}[t]{2.6in}
\begin{verbatim}
pmodels(S, PF) :-
    pmodels(S, PF, _).

:- table pmodels/3.
pmodels(S, until(SF1, SF2), H) :-
    models(SF2).
pmodels(S, until(SF1, SF2), H) :-
    models(SF1),
    trans(S, H, T),
    pmodels(T, until(SF1, SF2), next(H)).
pmodels(S, next(SF), H) :-
    trans(S, H, T),
    models(T, SF).

temporal(pmodels/3-3).
\end{verbatim}
\end{minipage}\\
\end{tabular}
\caption{Model checker for a fragment of PCTL}
\label{fig:pctl-mc}
\end{figure}

%\subsection{Generalized Probabilistic Logic (GPL)}
\paragraph{\textbf{GPL:}}  GPL is an expressive logic based on the modal mu-calculus for
probabilistic systems~\cite{GPL}.  GPL subsumes PCTL and PCTL* in
expressiveness. \comment{
We've introduced PCTL* now in related work
\footnote{PCTL is the sublogic of PCTL* in which every
  path formula is required to appear within the scope of a $\kw{pr}$
  operator.}
}
%\annotate{We haven't said what is PCTL*}.
GPL is designed for model checking \emph{reactive probabilistic transition
systems} (RPLTS), which are a generalization of DTMCs.  In an RPLTS,
a state may have zero or more outgoing transitions, each labeled by a
distinct action symbol.  Each action has a distribution on destination
states.  \comment{(In this way, an RPLTS is structurally similar to a Markov
Decision Process.)  A DTMC can be expressed as an RPLTS where each state
has at most one action, and all transitions are labeled by the same symbol.}

Syntactically, GPL has \emph{state} and \emph{fuzzy} formulae, where
the state formulae are similar to those of PCTL.  The fuzzy formulae are,
however, significantly more expressive.  The syntax of GPL, in
equational form, is given by:
\[
\begin{array}{rrl}
  \id{SF} &\Coloneqq & \kw{prop}(A)\ |\ \kw{neg}(\kw{prop}(A))\ |\ 
  \kw{and}(\id{SF}, \id{SF})\ |\ \kw{or}(\id{SF}, \id{SF}) \\
&  | & \kw{pr}(\id{PF}, \kw{gt}, B) 
\ |\  \kw{pr}(\id{PF}, \kw{lt}, B) 
\ |\  \kw{pr}(\id{PF}, \kw{geq}, B) 
\ |\  \kw{pr}(\id{PF}, \kw{leq}, B) \\
\id{PF} &\Coloneqq&  \kw{sf}(\id{SF})
\ |\ \kw{form}(X)
\ |\ \kw{and}(\id{PF}, \id{PF}) 
\ |\ \kw{or}(\id{PF}, \id{PF}) 
\ |\ \kw{diam}(A, \id{PF})
\ |\ \kw{box}(A, \id{PF})\\
\id{D} & \Coloneqq & \kw{def}(X, \kw{lfp}(\id{PF}))
\ |\  \kw{def}(X, \kw{gfp}(\id{PF}))
\end{array}
\]
Formula $\kw{diam}(A,\id{PF})$ holds at a state if there is an
$A$-transition after which \id{PF} holds; $\kw{box}(A, \id{PF})$ holds at
a state if \id{PF} holds after \emph{for every $A$-transition}.  
Least- and greatest-fixed-point formulas are written as a definition
$D$ using \texttt{lfp} and \texttt{gfp}, respectively.  
Formulae are specified as a set of definitions.  GPL admits only
alternation-free fixed-point formulae.  \comment{, and hence treating them as a set of
recursive definitions suffices to ensure the completeness of the
equational representation.}

\begin{figure}
\centering
\small
\begin{tabular}{ll}
  \multicolumn{2}{l}{\%\%\ \kw{pmodels(S, PF, H)}: \kw{S} is in the model of %
    fuzzy formula \kw{PF} at or after instant \kw{H}%
}\\
  \multicolumn{2}{l}{\%\%\ \kw{smodels(S, SF)}: \kw{S} is in the model of %
    state formula \kw{SF}%
}\\\hline
 \begin{minipage}[t]{2.5in} 
\begin{verbatim}
pmodels(S, sf(SF), H) :-
    smodels(S, SF).
pmodels(S, and(F1,F2), H) :-
    pmodels(S, F1, H),
    pmodels(S, F2, H).
pmodels(S, or(F1,F2), H) :-
    pmodels(S, F1, H);
    pmodels(S, F2, H).
pmodels(S, diam(A, F), H) :-
    trans(S, A, SW),
    msw(SW, H, T),
    pmodels(T, F, [T,SW|H]).
pmodels(S, box(A, F), H) :-
    findall(SW, trans(S,A,SW), L),
    all_pmodels(L, S, F, H).
\end{verbatim}
  \end{minipage}
&
  \begin{minipage}[t]{2.5in} 
\begin{verbatim}
pmodels(S, form(X), H) :-
    tabled_pmodels(S, X, H1), H=H1.

all_pmodels([], _, _, _H).
all_pmodels([SW|Rest], S, F, H) :-
    msw(SW, H, T),
    pmodels(T,F,[T,SW|H]),
    all_pmodels(Rest, S, F, H).

:- table tabled_pmodels/3.
tabled_pmodels(S,X,H) :-
    fdef(X, lfp(F)),
    pmodels(S, F, H).
\end{verbatim}
  \end{minipage}
\end{tabular}
\caption{Model checker for fuzzy formulas in GPL}
\label{fig:gpl-mc}
%\vspace*{-2em}
\end{figure}

A part of the model checker for GPL that deals with fuzzy formulae is
shown in Fig.~\ref{fig:gpl-mc}.  Note that fuzzy formulae have
probabilistic semantics, and, at the same time, may involve
conjunctions or disjunctions of other fuzzy formulae.  Thus, for
example, when evaluating $\kw{models}(s, \kw{and}(\id{PF}_1,
\id{PF}_2), H)$, the explanations of $\kw{models}(s, \id{PF}_1, H)$
and $\kw{models}(s, \id{PF}_2, H)$ may not be pairwise independent.
Thus recursion-free fuzzy formulae cannot be evaluated in PRISM, but
can be evaluated using the BDD-based algorithms of ProbLog and PITA.
In contrast,
\emph{recursive} fuzzy formulae can be evaluated using PIP.

\comment{
We can encode GPL analogously to PCTL, but finding the probability of
a fuzzy formula is more complicated; e.g., consider
$\textsf{Pr}(\psi_1 \OR \psi_2)$, the probability of the disjunction
holding at some state $s$.  If $\psi_1$ and $\psi_2$ were mutually
exclusive, then $\textsf{Pr}(\psi_1 \OR \psi_2) = \textsf{Pr}(\psi_1)
+ \textsf{Pr}(\psi_2)$.  Similarly, if they were independent,
$\textsf{Pr}(\psi_1 \AND \psi_2) = \textsf{Pr}(\psi_1) \cdot
\textsf{Pr}(\psi_2)$.  The PRISM implementation relies on such
properties.  If they do not hold, though, we must combine not the
respective probabilities of the two formulae, but their sets of
explanations.  ProbLog uses BDDs to accomplish that, and its
implementation would be able to evaluate recursion-free fuzzy
formulae.
}

\paragraph{\textbf{Recursive Markov Chains:}}
\begin{figure}[b]
  \centering\includegraphics[scale=.6]{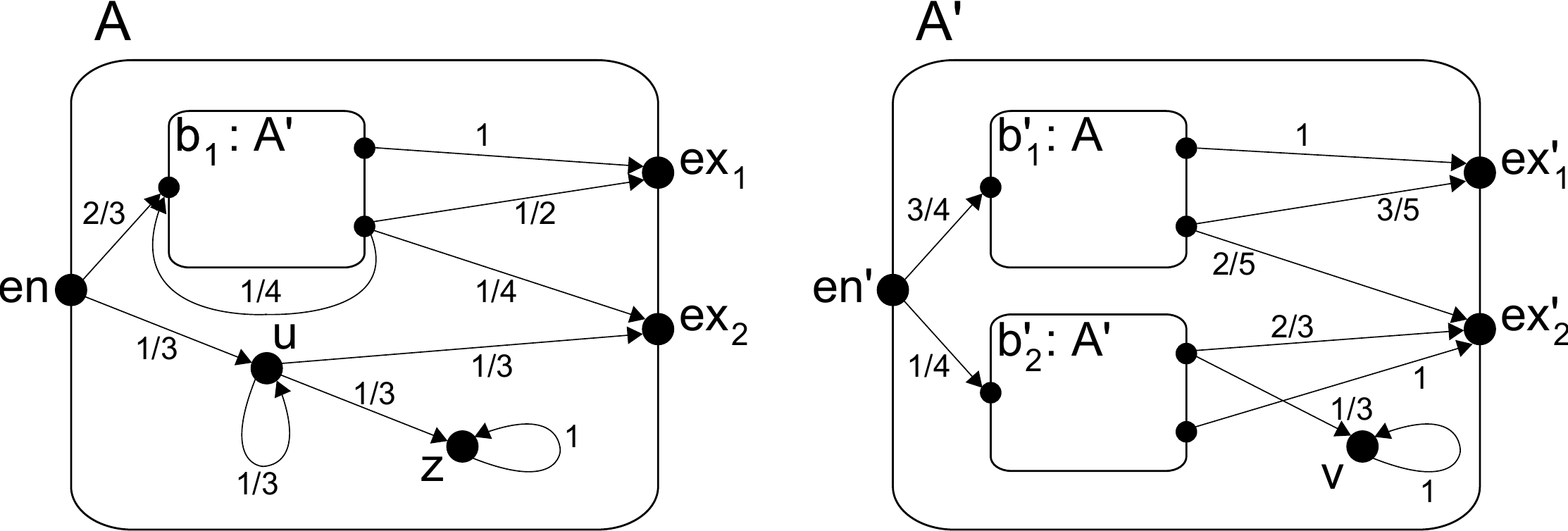}
  \caption{Example of a Recursive Markov Chain (RMC)}\label{fig:rmc}
\end{figure}

A Recursive Markov Chain (RMC) consists of \emph{components}, which
are analogous to procedure definitions in a procedural language.  The
structure of each component is similar to an automaton, with the
addition of \emph{boxes} that represent procedure calls.  An RMC can
be considered as an extension of DTMCs with recursively-called
components.  An example RMC from~\cite{RMC} is shown in
Fig.~\ref{fig:rmc}.
There are four special node types in an RMC: 
\emph{entry} ($en$) and \emph{exit nodes} ($ex$) associated with
components and \emph{call} and \emph{return ports} associated with
boxes.
In a box, call and return ports correspond to entry and exit nodes,
respectively, of the called component.  Behaviors of an RMC are the
set of runs with matching calls and returns.  Hence behaviors of an
RMC form a context free language.  We pose the problem of reachability
in an RMC (i.e. the probability of the set of runs that hit a given
state) in terms of GPL model checking of a corresponding RPLTS.

\begin{figure}
  \centering\includegraphics[scale=.6]{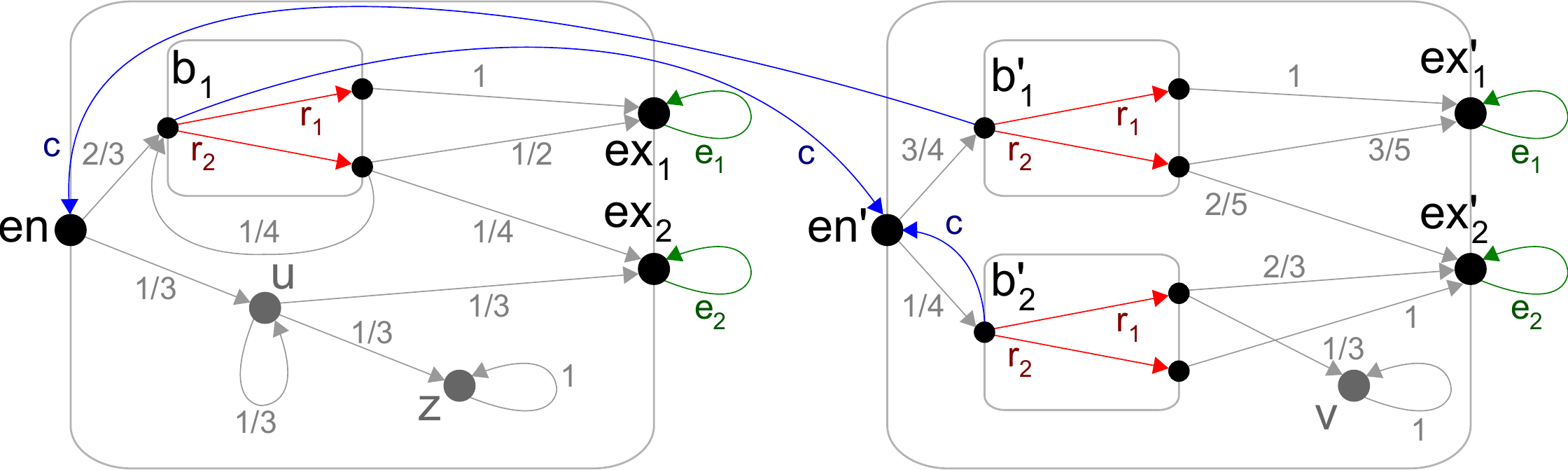}
  \caption{RPLTS corresponding to the example RMC in
    Fig.~\protect{\ref{fig:rmc}}}
\label{fig:rmctr}
\end{figure}

Given an RMC $R$, in which the \emph{maximum} number of exits in any
component is $n$, we define an RPLTS $R'$ and a set of $n$ mutually
recursive GPL formulae $X_1, X_2, \ldots, X_n$.  $R'$ will have a
state for every node of $R$, in particular including the call and
return ports of each box.  For each probabilistic transition in $R$,
we add the same transition in $R'$ and label it as $p$.  To model the
recursive call, we introduce three types of transitions.  From the
state $s$ in $R'$ corresponding to a call port in $R$, we add a $c$
(call) transition to state $s'$ corresponding to the called
component's entry node.  We also add $r_i$ (return) transitions from
state $s$ to the states corresponding to the return ports of the call.
Finally, from state $s$ in $R'$ corresponding to an exit node
$\id{ex}_i$, we add an $e_i$ (exit) transition back to $s$.  The exit
transitions and their labels enable us to write GPL formulae that
check for termination.  The RPLTS corresponding to the example RMC in
Fig.~\ref{fig:rmc} is shown in Fig.~\ref{fig:rmctr}.  Labels on
probabilistic transitions ($p$) are omitted in Fig.~\ref{fig:rmctr} to
avoid clutter.

We use GPL formulae $X_i$ to specify behaviors in an RMC that
eventually reach an exit node $\id{ex}_i$ of a component.  For a
2-exit RMC, the definitions of $X_i$ are as follows:
\begin{align*}
  \kw{def}(X_1, \kw{lfp}(\kw{or}(&\kw{diam}(e_1, \kw{tt}),
  \\ \kw{or}(&\kw{diam}(p, X_1),
  \\ \kw{or}(&\kw{and}(\kw{diam}(c, X_1), \kw{diam}(r_1, X_1)),
  \\ &\kw{and}(\kw{diam}(c, X_2), \kw{diam}(r_2, X_1))))))).\\
  \kw{def}(X_2, \kw{lfp}(\kw{or}(&\kw{diam}(e_2, \kw{tt}),
  \\ \kw{or}(&\kw{diam}(p, X_2),
  \\ \kw{or}(&\kw{and}(\kw{diam}(c, X_1), \kw{diam}(r_1, X_2)),
  \\ &\kw{and}(\kw{diam}(c, X_2), \kw{diam}(r_2, X_2))))))).
\end{align*}
The intuition behind the formulae for $X_i$ is as follows.  Since
$X_i$ specifies that $\id{ex}_i$ is eventually reached, $X_i$ is a
least fixed point formula.  The ways in which exit $\id{ex}_i$ is
reached from a state $s$ are:
\begin{enumerate}
\item  $e_i$ transition is enabled at $s$:  this corresponds to the
  disjunct $\kw{diam}(e_i,\kw{tt})$ in the definition of $X_i$;
\item there is a probabilistic transition ($p$) from $s$ to $s'$ such
  that $\id{ex}_i$ is reached eventually from $s'$ (corresponds to
  ($\kw{diam}(p,X_i)$);
\item $s$ corresponds to a call, that call eventually returns from
  $\id{ex}_j$ for some $j$, and subsequently, $\id{ex}_i$ is reached.
  The formula $\kw{and}(\kw{diam}(c, X_j),\kw{diam}(r_j, X_i))$
  encodes this way of reaching $\id{ex}_i$ via $\id{ex}_j$.  The
  subformula $\kw{diam}(c, X_j)$ specifies that $\id{ex}_j$ is reached
  after the call; and the subformula $\kw{diam}(r_j, X_i)$ specifies
  that after the corresponding return, $\id{ex}_i$ is eventually
  reached.
\end{enumerate}

While the example shows the GPL formulae for 2-exit RMCs, the
description above gives the general structure of the formulae for
$n$-exit RMCs.  Note that if a component has fewer than $n$ exits,
then the formula $X_n$ will be trivially false at all of its nodes.
Moreover, behaviors satisfying $X_i$ and those satisfying $X_j$ ($i
\not = j$) are mutually exclusive, since we cannot terminate at more
than one exit on a single path.  Finally, the GPL formula is the same
regardless of what RMC we are attempting to transform, and only
depends on $n$.

\comment{
Recursive Markov Chains (RMCs) use special \emph{call} and
\emph{return} nodes to provide a model of DTMCs with a
recursive-call capability.  We can check reachability in an RMC
by generating a corresponding RPLTS, and model checking a standard GPL
formula w.r.t. that RPLTS.
See \cite{techreport} for the details.
%I think we need not give details of RMCs in this paper;  we can have
%a paragraph or so describing how they can be model checked within
%our framework, but we do not need to give formal details or even an
%example!.
}

\section{Experimental Results}
\label{sec:results}

%\subsection{Prototype Implementation}
\newalg has been implemented using the XSB tabled logic programming
system~\cite{XSB}.  An explanation generator is constructed by
performing normal query evaluation under the well-founded semantics by
redefining \texttt{msw}s to backtrack through their potential values,
and have the \emph{undefined} truth value.  This generates a
\emph{residual} program in XSB that captures the dependencies between
the original goal and the \texttt{msw}s (now treated as undefined
values).  In one partial implementation, called \textbf{\newalg-Prism},
the probabilities are computed directly from the residual program.
Note that such a computation will be correct if PRISM's restrictions
are satisfied.  In general, however, we materialize the residual
program  into a dynamic database that corresponds
to the productions in the explanation generator.  A second partial
implementation, called \textbf{\newalg-BDD}, constructs BDDs from the
explanation generator, and computes probabilities from the BDD.  Note
that \textbf{\newalg-BDD} will be correct when the finiteness restriction
holds.  The full implementation of \newalg, called
\textbf{\newalg-full}, is obtained by constructing a set of FEDs from
the explanation generator (Def.~\ref{def:fed-construction}),
generating polynomial equations from the set of FEDs
(Def.~\ref{def:ff-to-eqn}) and finally finding the least solution to
the set of equations.  The final equation solver is implemented in C.
All other parts of the three implementations,
including the BDD and FED structures, are completely implemented in
tabled Prolog.

%\subsection{Experimental Results}
We present two sets of experimental results, evaluating the performance of
\newalg on
(1)~programs satisfying PRISM's restrictions;  and
(2)~a program for model checking PCTL formulae.

\subparagraph{Performance on PRISM Programs:}
Note that all three implementations--- \newalg-Prism, \newalg-BDD and
\newalg-full may be used to evaluate PRISM programs.  

\begin{figure}
  \centering
  \begin{tabular}{c@{\extracolsep{2em}}c}
    \begin{minipage}{1.75in}
     \includegraphics[scale=0.45]{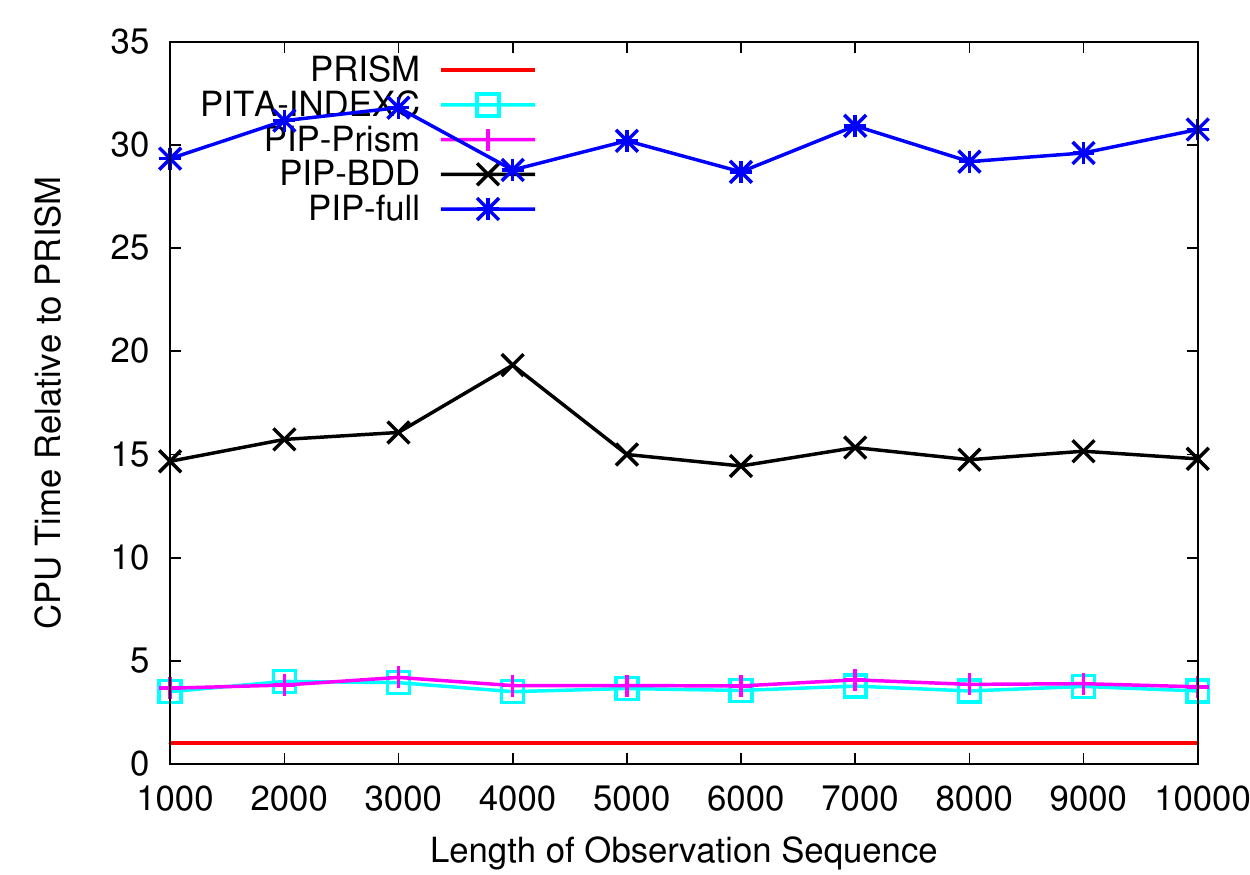}
    \end{minipage}
&
    \begin{minipage}{2in}
     \includegraphics[scale=0.45]{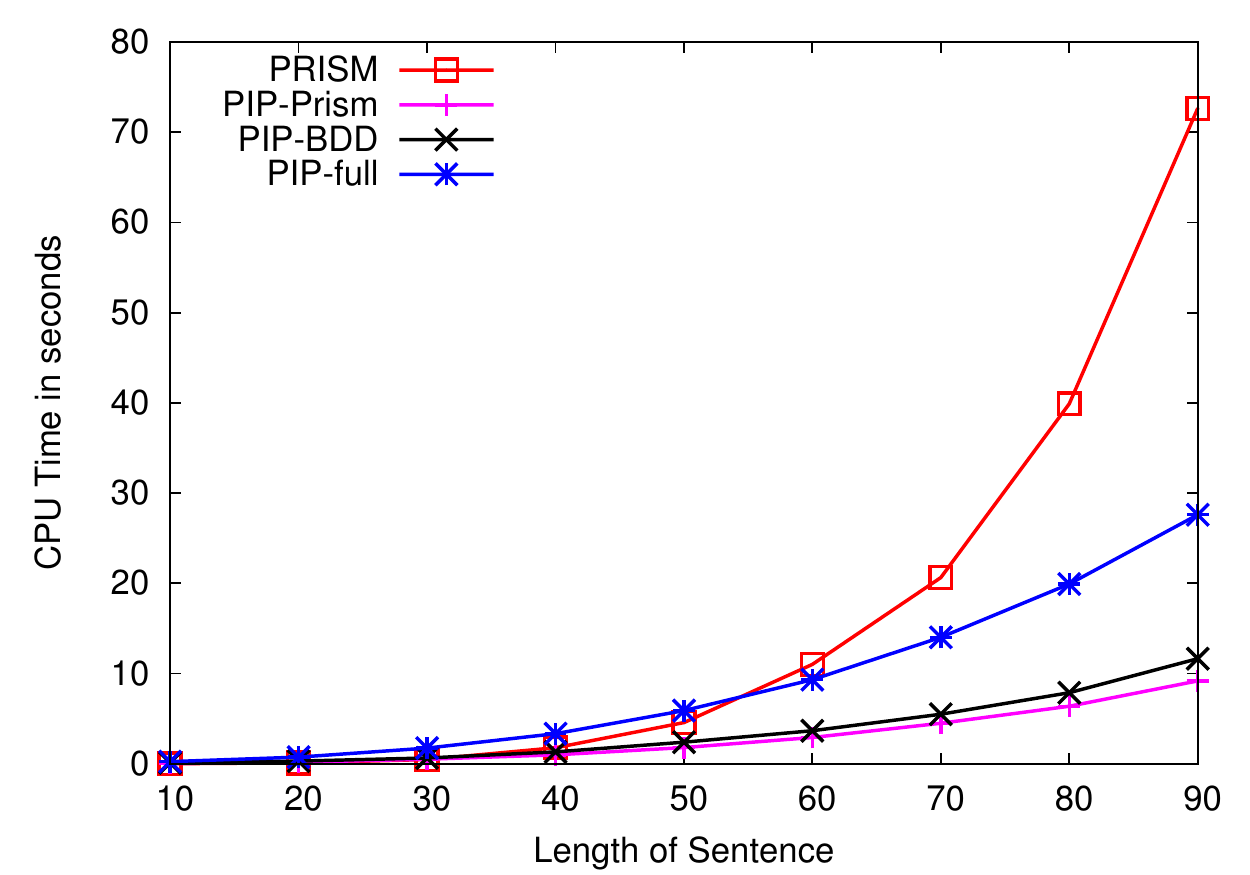}
    \end{minipage}\\
(a) Relative performance on HMM & (b) Performance for NCN queries for PLC\\
    \begin{minipage}{1.75in}
     \includegraphics[scale=0.45]{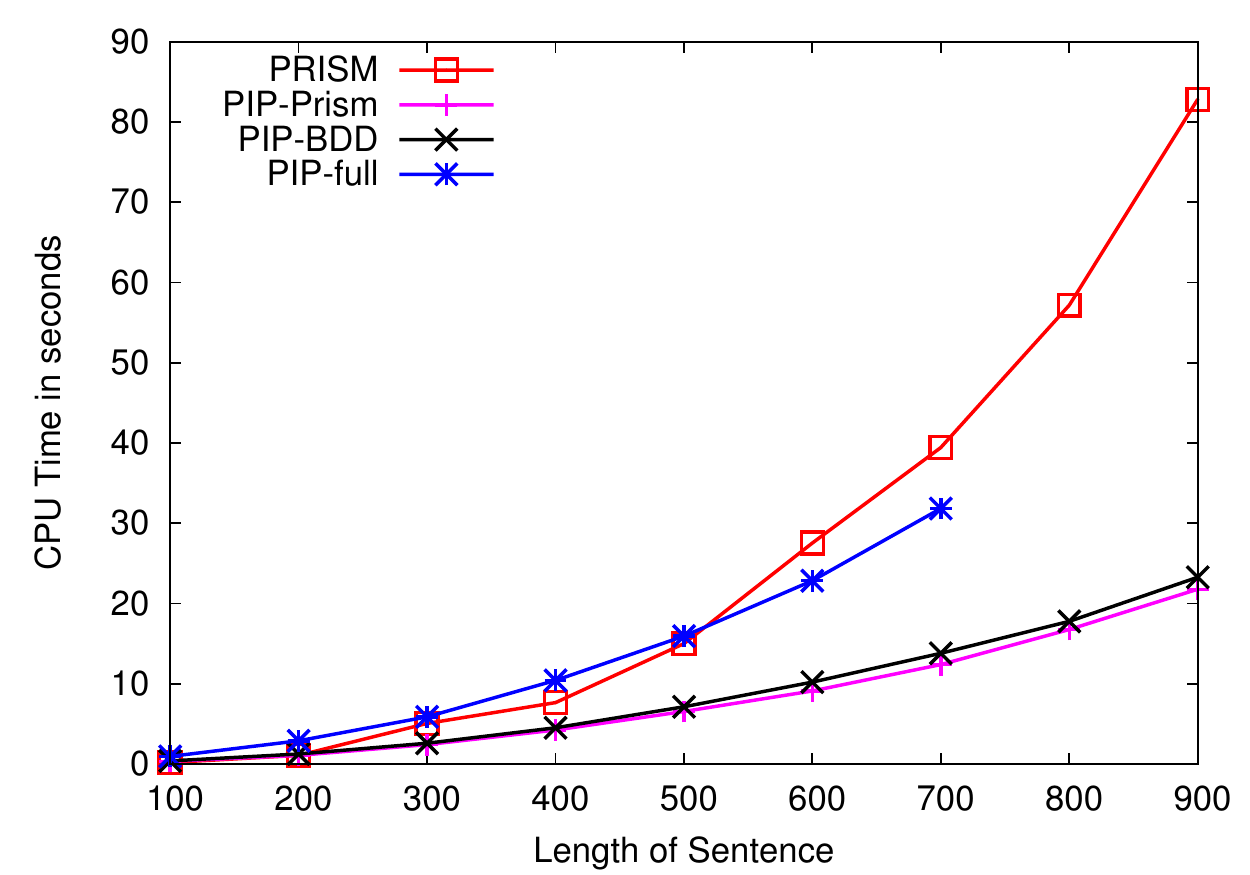}
    \end{minipage}
&
    \begin{minipage}{2in}
     \includegraphics[scale=0.45]{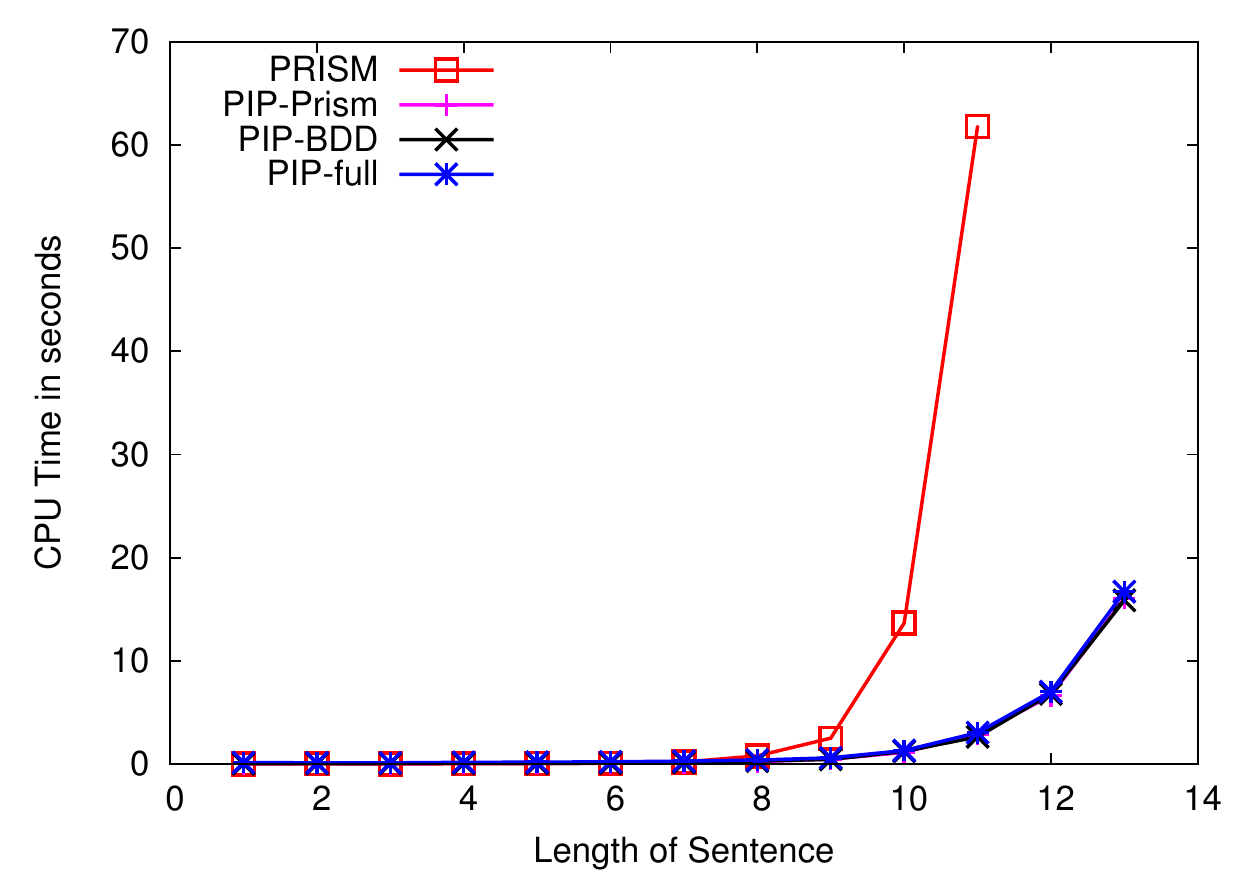}
    \end{minipage}\\
(c) Performance for NPV queries for PLC & (d) Performance for ADVN queries for PLC\\
 \end{tabular}
  \caption{Performance of \newalg on PRISM Programs}
  \label{fig:prism-results}
\end{figure}

\paragraph{Hidden Markov Model (HMM):}
We used the simple 2-state gene sequence HMM from~\cite{CG:2009} (also used
in~\cite{RiguzziSwift11}) for our evaluation.  
We measured the CPU time taken by the versions of
\newalg, PRISM 2.0.3 and PITA-INDEXC~\cite{RiguzziSwift11} (a version
of PITA that does not use BDDs and uses PRISM's assumption) to evaluate the
probability of a given observation sequence, for varying sequence lengths.
The observation sequence itself was embedded as a set
of facts (instead of an argument list).  This makes table accesses
fast even when shallow indices are used.  The performance of the three
\newalg versions and PITA-INDEXC, relative to PRISM~2.0.3, is shown in
Fig.~\ref{fig:prism-results}(a).  CPU times are
normalized using PRISM's time as the baseline.  Observe
that the \newalg-Prism and PITA-INDEXC perform similarly: about 3.5 to 4
times slower than PRISM.  Construction of BDDs (done in \newalg-BDD, but
not in \newalg-Prism) adds an extra  factor of 4 overhead.  Construction
of full-fledged FEDs, generating polynomial equations and solving them
(done only in \newalg-full) adds another factor of 2 overhead.  We find
that the equation-solving time (using the only component coded in C)
is generally negligible.

\paragraph{Probabilistic Left Corner Parsing:}
This example was adapted from PRISM's example suite, parameterizing
the length of the input sequence to be parsed.  We measured the CPU
time taken by the three versions of \newalg and PRISM~2.0.3 on a machine with an
Intel Pentium 2.16GHz processor.  The performance on
three queries (each encoding a different class
of strings) is shown in Fig.~\ref{fig:prism-results}(b)--(d).  In contrast to
the HMM example, the sequences are represented as lists.  For these
examples, \newalg-Prism implementation outperforms PRISM.  Moreover,
although \newalg-BDD and \newalg-full are slower than \newalg-Prism, the
relative performance gap is much smaller than observed in the HMM
example.

\begin{figure}
  \centering
  \begin{tabular}{c@{\extracolsep{6em}}c}
    \begin{minipage}{1.75in}
     \includegraphics[scale=0.45]{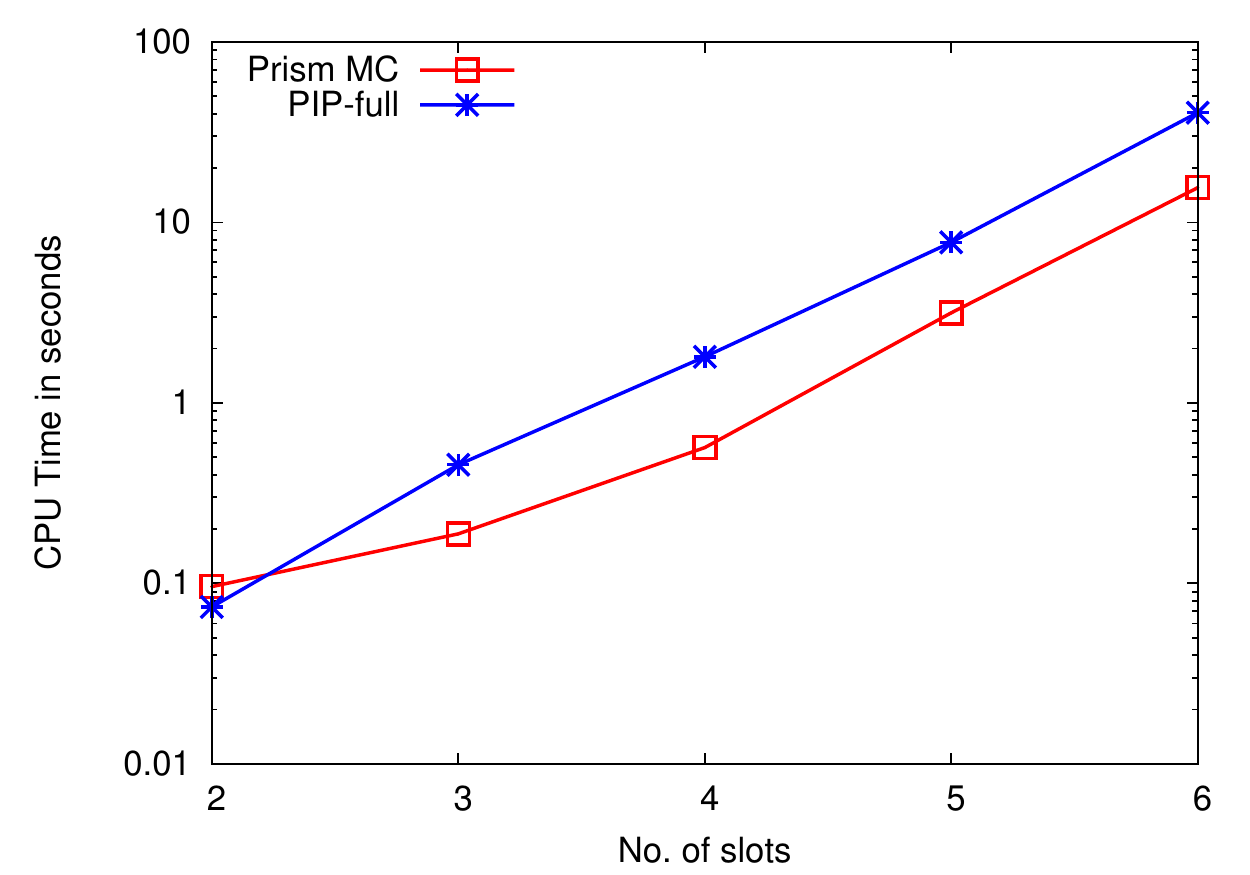}
    \end{minipage}
&
    \begin{minipage}{2in}
     \includegraphics[scale=0.45]{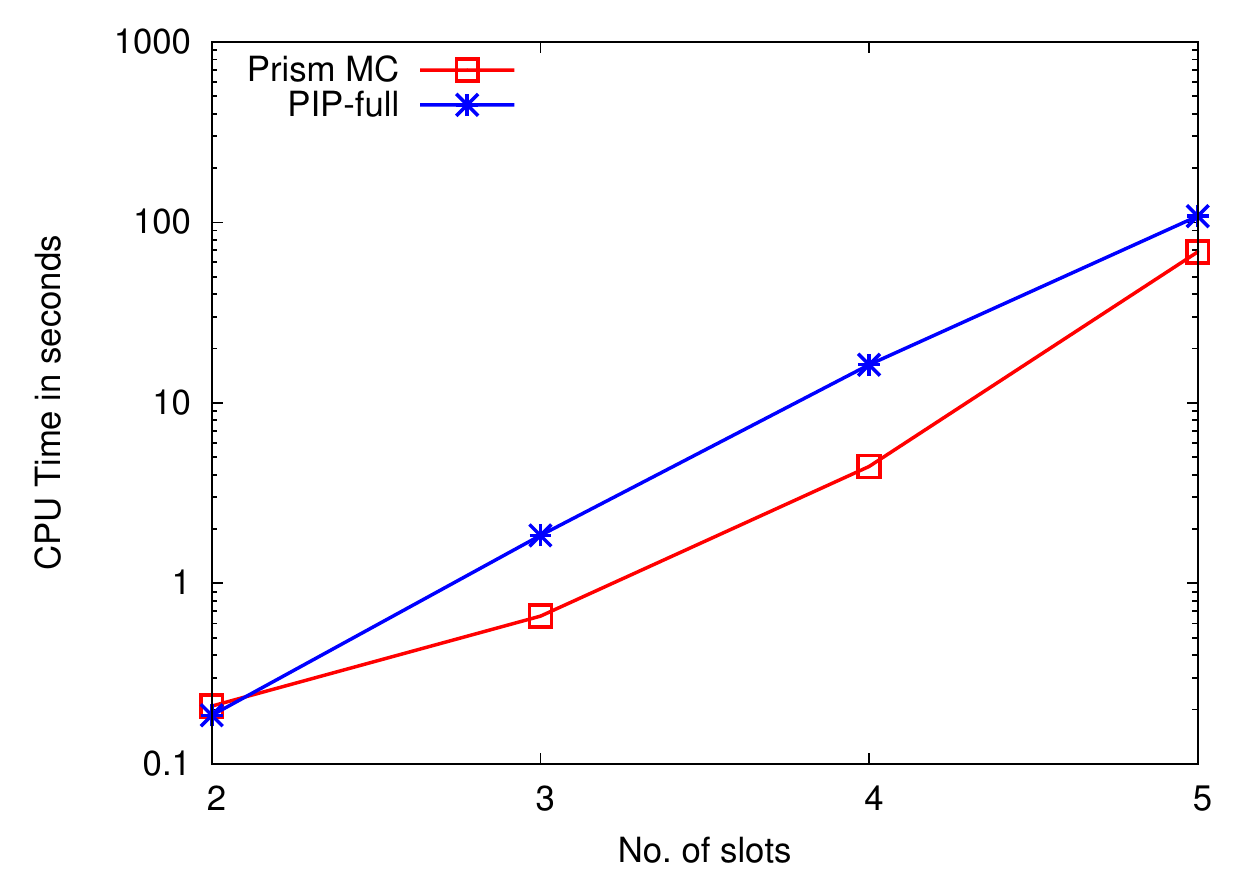}
    \end{minipage}\\
(a) $N= 5$ & (b) $N = 6$
 \end{tabular}
 \caption{Performace of PCTL model checking using \newalg and the
   Prism model checker for Synchronous Leader Election protocol of different sizes}
  \label{fig:mc-results}
%\vspace*{-2em}
\end{figure}

\subparagraph{Performance of the PCTL Model Checker:}
We evaluated the performance of \newalg-full for supporting a PCTL model
checker (encoded as shown in Fig.~\ref{fig:pctl-mc}).  We compared the
performance of \newalg-based model checker with that of the widely-used
Prism model checker~\cite{PRISMMC}.  We show the
performance of PIP and the Prism model checker on the Synchronous Leader
Election Protocol~\cite{ItaiRodeh81} for 
computing the probability that eventually a leader will be elected.
Fig.~\ref{fig:mc-results}
shows the CPU time used to compute the
probabilities of this property on systems of different sizes.  Observe that our
high-level implementation of a model checker based on \newalg performs
within a factor of 3 of the Prism model checker (note: the y-axis on
these graphs is logarithmic).  Moreover, the two
model checkers show similar performance trends with increasing problem
instances.  However, it should be noted
that the Prism model checker uses a 
BDD-based representation of reachable states, which
can, in principle, scale better to large state spaces compared to the
explicit state representation used in our \newalg-based model checker.

\section{Conclusions}
\label{sec:conc}

In this paper, we have shown that in order to formulate the problem of
probabilistic model checking in probabilistic logic programming, one
needs an inference algorithm that functions correctly even when
finiteness, mutual-exclusion, and independence assumptions are
simultaneously violated.  We have presented such an inference algorithm,
\newalg, implemented it in XSB Prolog, and demonstrated its practical
utility by using it as the basis for encoding model checkers for a rich
class of probabilistic models and temporal logics.

For future work, we plan to refine and strengthen the implementation
of \newalg.  We also plan to explore more substantial model-checking
case studies.  It would be interesting to study whether
optimizations to exploit data independence and symmetry, 
which are easily enabled by high-level encodings of model checkers, will be 
effective for probabilistic systems as well.

\paragraph{Acknowledgments.}
Research supported in part by 
NSF Grants CCF-\mbox{1018459} % Probabilistic Tabled Logic Programming
CCF-\mbox{0926190}, % CMACS
CCF-\mbox{0831298}, % CT-T
AFOSR Grant FA9550-09-1-0481,  % Survivable Software
and ONR Grant N00014-07-1-0928. % Framework for Analyzing & Ensuring Trust

\end{document}